\documentclass[aps,
               reprint,
               prb,
               twocolumn,
               superscriptaddress]{revtex4-2}
\usepackage{bm}
\usepackage{placeins}
\usepackage{graphicx}
\usepackage{graphics}
\usepackage{amsmath}
\usepackage{amsfonts}
\usepackage{amssymb}
\usepackage{makeidx}
\usepackage[colorlinks=true,citecolor=blue,linkcolor=blue,linktocpage=true,pagebackref=false]{hyperref}
\hypersetup{colorlinks=true,citecolor=blue,linkcolor=blue,filecolor=blue,urlcolor=blue}
\usepackage{color,soul} 
\usepackage{float}
\usepackage{subfigure}
\usepackage{physics}
\usepackage{xcolor}
\usepackage[normalem]{ulem}

\usepackage{siunitx}

\newcommand{\orcid}[1]{\href{https://orcid.org/#1}{\includegraphics[width=8pt]{orcid.pdf}}}
\graphicspath{{figures/}} 
\begin{document}

\newcommand{\be}   {\begin{equation}}
\newcommand{\ee}   {\end{equation}}
\newcommand{\ba}   {\begin{eqnarray}}
\newcommand{\ea}   {\end{eqnarray}}
\newcommand{\ve}   {\varepsilon}
\newcommand{\Dis}  {\mbox{\scriptsize dis}}

\newcommand{\state} {\mbox{\scriptsize state}}
\newcommand{\band} {\mbox{\scriptsize band}}


\title{Unveiling the Electronic Origin of Anomalous Contact Conductance in \\ Twisted Bilayer Graphene}

\author{Kevin J. U. Vidarte}
\affiliation{Instituto de F\'{\i}sica, Universidade Federal do Rio de Janeiro, 
             21941-972 Rio de Janeiro, RJ, Brazil}

\author{Caio Lewenkopf 
}
\affiliation{Instituto de F\'{\i}sica, Universidade Federal do Rio de Janeiro, 
             21941-972 Rio de Janeiro, RJ, Brazil}
             
\author{F. Crasto de Lima 
} 
\affiliation{Ilum School of Science, Brazilian Center for Research in Energy and Materials (CNPEM), Campinas, SP, Brazil}

\author{R. Hiroki Miwa}
\affiliation{Instituto de Física, Universidade Federal de Uberlândia, 38400-902, Uberlândia, MG, Brazil}

\author{Felipe Pérez Riffo}
\affiliation{Departamento de Física, Universidad Técnica Federico Santa María, Casilla 110-V, Valparaíso, Chile}

\author{Eric Suárez Morell}
\affiliation{Departamento de Física, Universidad Técnica Federico Santa María, Casilla 110-V, Valparaíso, Chile}

\date{\today}
\begin{abstract}
This study theoretically investigates the contact conductance in twisted bilayer graphene (TBG),  {addressing} recent experimental observations from scanning tunneling microscopy (STM) and conductive atomic force microscopy (c-AFM).
These experiments revealed a surprising non-monotonic current pattern as a function of the TBG rotation angle $\theta$, with a peak at $\theta \approx 5^\circ$, a finding that markedly departs from the well-known magic angle TBG behavior.
To elucidate this phenomenon, we develop a theoretical and computational framework  {based on the system's local density of states, and evaluate the contact current over the energy range of states that contribute to the junction}.
{Our results, obtained for both relaxed and rigid TBG structures, indicate that the conductance maximum at $\theta \approx 5^{\circ}$ is not primarily caused by structural relaxation or by changes in the AA-stacking area: the maximum persists, at essentially the same angle, in rigid (unreconstructed) lattices.}
Instead, we attribute this peak to the evolution of the electronic band structure, specifically the shifting of van Hove singularities (vHs) toward the Fermi level as the twist angle decreases.
 {The precise location of the maximum tracks the energy window of states participating in the contact, which in contact-mode measurements is set by the local doping under the metallic tip rather than by the small applied bias alone.
We show that the angle of maximum conductance shifts with carrier doping, a dependence that is directly testable by electrostatic gating and that a purely structural mechanism cannot produce.}
\end{abstract}
\maketitle

\section{Introduction}
\label{sec:introduction}

Twisted bilayer graphene (TBG) systems offer a new platform for exploring emergent quantum phenomena in condensed matter\cite{Andrei2020}.
By controlling the twist angle between two graphene layers, one can induce a periodic moiré superlattice that dramatically alters the electronic band structure \cite{LopesdosSantos2007}. 
At specific ``magic angles", flat bands appear near the Fermi level \cite{Morell2010, Laissardiere2010, Bistritzer2011}, enhancing electron-electron interactions and enabling the formation of correlated insulating and superconducting phases \cite{Cao2018a, Cao2018b, Yankowitz2019}. 
These findings have spurred significant interest in understanding the intricate interplay between topology and interactions in TBG systems.

Important insights on the properties of TBG systems can be obtained from scanning probe microscopy (STM) \cite{Wong2015, Tilak2021, Calugaru2022, Brihuega2012} and standard and conductive atomic force microscopy (c-AFM) \cite{Koren2016, Chari2016, Liao2018, Yu2020,  Zhang2020, Huang2021}. 
By measuring the contact conductivity of TBG systems as a function of the twist angle $\theta$, these techniques provide valuable quantitative information on the system's local density of states (LDOS) and lattice relaxation.

Bistritzer and MacDonald predicted six interlayer conductivity peaks in twisted few-layer graphene at specific twist angles, including $\theta=21.8^{\circ}$ and $38.2^{\circ}$ \cite{Bistritzer2010}. 
Experimental evidence supporting these predictions was provided by two independent groups \cite{Koren2016,Chari2016} using conductive Atomic Force Microscopy (c-AFM).
These studies observed enhanced interlayer conductivity at the aforementioned angles in graphite-based systems.
While these findings partially validate the theoretical predictions, the absence of the other four predicted peaks remained a puzzle. 

Early experimental work \cite{Kim2013} demonstrated that incommensurate interfaces, lacking Fermi surface overlap, suppress interlayer coupling and conductivity. 
Subsequent studies experimentally investigated the dependence of the interlayer conductivity on twist-angle $\theta$ in   {TBG} systems  \cite{Yu2020, Zhang2020} and even in other Van der Waals materials, such as MoS$_2$/graphene heterostructures \cite{Liao2018}.
Ref.~\cite{Yu2020} did not observe an exceptionally enhanced conductivity at the angles predicted by theory \cite{Bistritzer2010} and found a linear dependence on the $\theta$ for large twist-angles, consistent with the theoretical prediction based on a phonon-mediated interlayer transport mechanism, considering the renormalization of the Fermi velocity \cite{Perebeinos2012,Perebeinos2024}.
Zhang and collaborators \cite{Zhang2020} reported a maximum in interlayer conductivity around $\theta \approx 5^\circ$ followed by a decrease in larger twist-angle TBG samples.
This behavior was attributed to the increasing dominance of low-conductivity AB-stacked regions in the moiré pattern. 
However, their interpretation based on {\it ab initio} carrier density may not be entirely accurate, as the energy window used for integration is somewhat too large to directly correlate with STM measurements.
Despite significant progress, a comprehensive understanding of vertical transport mechanisms in bilayer or few-layer two-dimensional (2D) electronic systems remains elusive \cite{Liao2018, Zhang2020, Yu2020}.


In this study, we investigate the contact conductance in TBG systems, specifically addressing recent experimental observations from scanning tunneling microscopy (STM). 
Our approach involves simulating the system's lattice reconstruction and computing its local density of states (LDOS).
Contrary to previous predictions, our findings reveal that the experimentally observed maximum in local conductance at $\theta \approx 5^\circ$ is not a consequence of structural relaxation. 
Instead, we demonstrate that this peak is directly linked to the evolution of the electronic band structure as a function of the twist angle, particularly the shifting of van Hove singularities (vHs) to the Fermi level. 
Furthermore, we show that the contact conductance remains practically the same for both relaxed and rigid TBG systems at a given twist angle.

This paper is organized as follows. 
In Sec.~\ref{sec:method_and_methods}, we discuss the description of TBG crystal structures as a function of the twist angle $\theta$, including their symmetries, and outline our approach for computing lattice relaxation effects. 
We also introduce an efficient methodology to estimate the size of the AA-stacking region.
Next, we present the effective tight-binding Hamiltonian employed to calculate the low-energy electronic properties of TBGs and detail our simulation approach for STM measurements. 
In Sec.~\ref{sec:results}, we quantify the in-plane and out-of-plane lattice relaxation effects as a function of the twist angle $\theta$ and compare the LDOS for each case. 
Later, we show how the peak in the local conductance depends on the relative rotational angle and on the STM bias. 
We discuss our results and present our conclusions in Sec.~\ref{sec:conclusion}.

\section{Theory and methods}
\label{sec:method_and_methods}

\subsection{Structural calculations}
\label{sec:structural-calculations}

TBG lattices are characterized, as standard \cite{LopesdosSantos2007, Rozhkov2016}, by the twist of one graphene layer with respect to the other around a given site starting from the AB-stacked bilayer, forming a moir\'e pattern.
We take the primitive lattice vectors of the bottom layer as $\textbf{a}^{b}_{1}=\sqrt{3}a_0/2 \left( \sqrt{3} \hat{\bf e}_{x} + \hat{\bf e}_{y} \right)$ and $\textbf{a}^{b}_{2}=\sqrt{3}a_0/2 \left( \sqrt{3}\hat{\bf e}_{x} - \hat{\bf e}_{y} \right)$ and those of top layer as $\textbf{a}^{t}_{i}=R(\theta)\textbf{a}^{b}_{i}$, where $R(\theta)$ is the rotation matrix.
For non-relaxed TBGs lattices, we set the graphene carbon-carbon bond length as $a_{0}=1.42$~\AA\,(with corresponding single-layer lattice constant $a = \sqrt{3} a_0)$, and the spacing between graphene layers as $d_{0} = 3.35$~\AA~\cite{Laissardiere2010}. 

The lattice structure of a TBG system is periodic if the periods of the two graphene layers match, giving a finite unit cell. 
Hence, the periodicity condition requires the lattice translation vector $m\textbf{a}_{1}^{b} + n \textbf{a}_{2}^{b}$ of the bottom (unrotated) layer and $n\textbf{a}_{1}^{t} + m\textbf{a}_{2}^{t}$ in the top (rotated) layer, with $m$ and $n$ integers, to coincide. 
The twist angle for a commensurate structure is related to $(m,n)$ by \cite{Moon2013,Mele2010,Mele2012} 
\be
\theta (m,n)= 	\arccos{\left( \dfrac{1}{2}\dfrac{m^{2}+n^{2}+4mn}{m^{2}+n^{2}+mn} \right) } ,
\ee
with a lattice constant
\be
\label{Eq:lattice_constant}
L_{m}=a_{0}\sqrt{3(m^{2}+n^{2}+mn)}=\dfrac{\vert m-n\vert\sqrt{3}a_{0}}{2\sin{\theta/2}},
\ee
which contains $N=4(m^{2}+n^{2}+mn)$ atoms.

\begin{figure}[h!]
    \centering
    \includegraphics[width=0.80\linewidth]{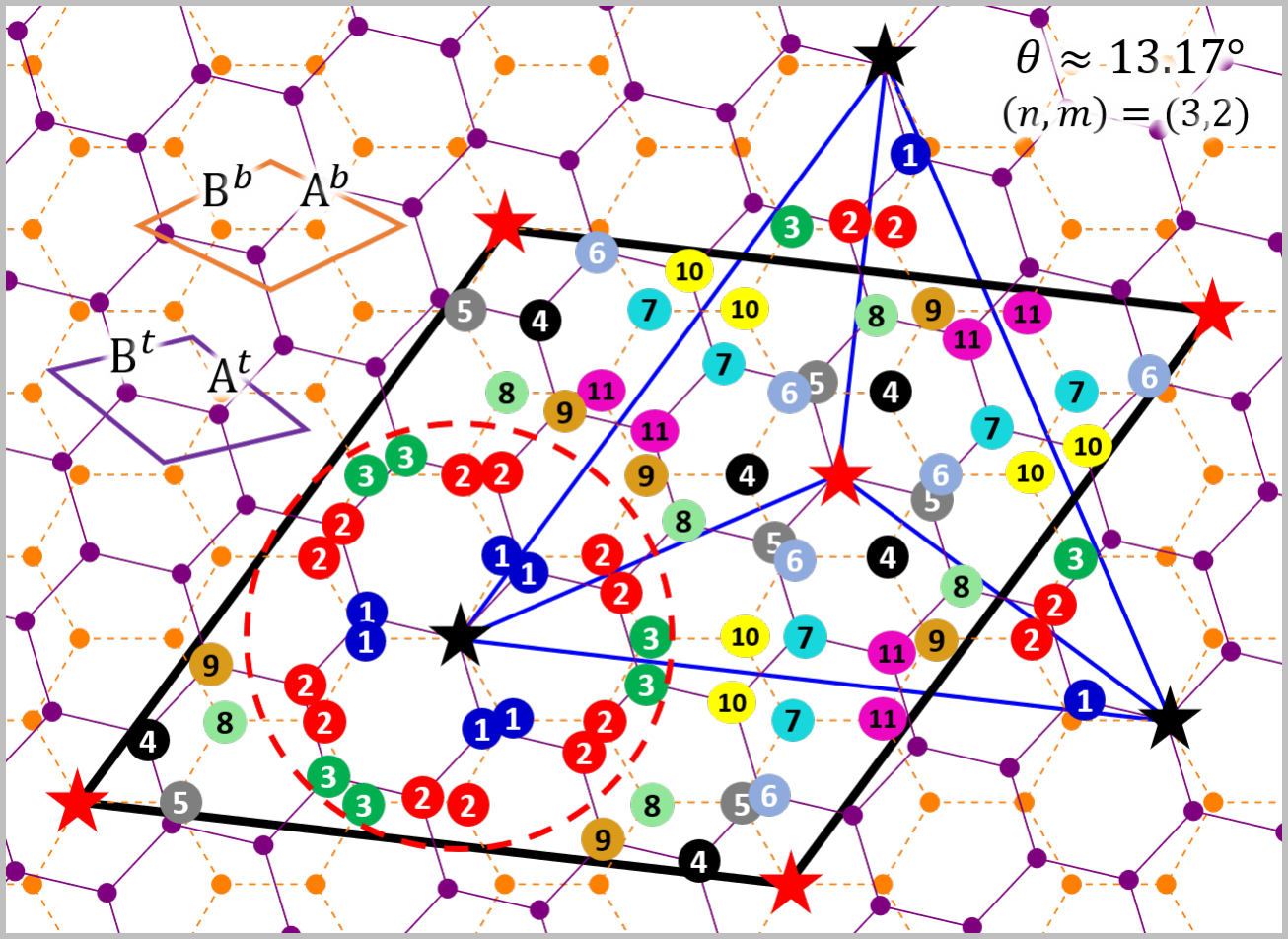}
    \caption{TBG commensurate structure with $\theta\approx13.17^{\circ}$ that corresponds to $(m,n)= (3,2)$.
    Purple (orange) circles and lines indicate the carbon sites and the primitive unit cell of the top (bottom) layer, respectively. 
    The black lines outline the primitive unit cell of the TBG system.
    Black (red) stars indicate the ${\rm A}^{b}{\rm A}^{t}$ (${\rm B}^{b}{\rm H}^{t}$ or ${\rm H}^{b}{\rm B}^{t}$) sites.
    Sites labeled by the same number (same color) are equivalent.
    The dashed red circle outlines the AA-stacking region.
    }
    \label{fig:EquivalentSites}
\end{figure}

Each graphene layer comprises two sublattices, denoted as ${\rm A}^{b({\rm or}\; t)}$ and ${\rm B}^{b({\rm or}\; t)}$.
Commensurate structures occur in two distinct forms identified by their sublattice parities \cite{Vidarte2025,Mele2010,Mele2012}.
Figure~\ref{fig:EquivalentSites} shows an example of an odd commensurate structure, which is characterized by having 3 three-fold symmetric positions, corresponding to the stacking of ${\rm B}^{b}{\rm H}^{t}$, ${\rm A}^{b}{\rm A}^{t}$ and ${\rm H}^{b}{\rm B}^{t}$ sites, indicated by star points.
We label by ${\rm B}^{b}{\rm H}^{t}$ (or ${\rm H}^{b}{\rm B}^{t}$) the sites where a ${\rm B}^{b}$ (or ${\rm B}^{t}$)-sublattice site of one layer aligns with the hexagonal centers ${\rm H}^{t}$ (or ${\rm H}^{b}$) of its neighboring layer.
We chose to investigate odd commensurate structures for convenience, but the parity of the system does not affect our findings.

To facilitate the analysis of TBG lattice structures and efficiently identify all distinct equivalent sites within a unit cell, let us examine their symmetries. 
In Fig.~\ref{fig:EquivalentSites}, the carbon atom sites marked by identical colored disks are equivalent due to their $C_{3}$ three-fold symmetric positions.
A practical approach to select the nonequivalent atomic sites is to connect (blue lines) the points of two ${\rm A}^{b}{\rm A}^{t}$ sites and one ${\rm B}^{b}{\rm H}^{t}$ (or ${\rm H}^{b}{\rm B}^{t}$) site. 
These lines form a triangular area encompassing one-sixth of the unit cell's total area.
Notably, the blue lines connecting two ${\rm A}^{b}{\rm A}^{t}$ sites correspond to the $C'_{2}$ two-fold rotation axes, providing a clear and efficient means to distinguish between all different equivalent sites within the unit cell.

Starting from the above-defined lattice structures, we determine the atomic relaxation positions for a given twist-angle $\theta$ using the Large-scale Atomic/Molecular Massively Parallel Simulator (LAMMPS) \cite{lammps_1995}. 
The energy minimization process incorporates both non-bonded and bonded interactions to model the system accurately.
Non-bonded interlayer interactions are handled using a registry-dependent interlayer potential (ILP) \cite{Leven2014,Leven2016,Maaravi2017}, with a cutoff distance of 16 \AA, ensuring that all relevant interlayer forces are considered. 
 {
We employ the refined ILP parametrization of Ouyang {\it et al.}~\cite{Ouyang2018}, fitted and validated for graphitic and h-BN homo- and heterostructures, including graphene bilayers. The structural corrugation obtained with this parametrization is in very good agreement with published benchmark DFT-quality full-relaxation results~\cite{Uchida2014,nam_lattice_2017}.
} 
Bonded intralayer interactions within each layer are modeled using the Tersoff potential \cite{Tersoff1988,Kinaci2012}, which is well-suited for describing the covalent bonding in carbon-based materials like graphene.  
The lattice structure relaxation calculations are performed with an energy tolerance criterion set to $10^{-10}$ eV per atom \cite{Mostofi_2020}, ensuring high precision in determining the equilibrium atomic positions.

\subsection{Evaluation of the AA-stacking region area}
\label{sec:AA-region-area}

Although the relaxation of the unit cell in rotated layers of van der Waals-type materials is a well-established topic \cite{nam_lattice_2017}, the assessment of the AA zone size depends on the cutoff criteria and/or the geometric method employed. We propose a simple mechanism based on the symmetry of atomic positions around the AA site to define the AA region, allowing for a direct comparison of relaxed and unrelaxed structures without resorting to arbitrary cuts.

In rigid lattice structures, the area of the AA-stacking region encompasses a set of equivalent and equidistant sites relative to an ${\rm A}^{b}{\rm A}^{t}$ site.
As shown in Fig.~\ref{fig:EquivalentSites}, for a twist angle of $\theta\approx 13.17^{\circ}$, only sites 1, 2, and 3 (depicted as blue, red, and green disks) are both equivalent and equidistant from the central ``black star" lattice site.
The radius of the AA-region area, $r^{\rm rig}_{\rm AA}\approx 2.85$~\AA, is defined by the furthest of these equivalent sites within the AA zone.
Notably, these sites belong to different layers, which are vertically aligned when $\theta = 0$. 
We denote their distance as $d_{c}$.
Conversely, the equidistant sites relative to a ${\rm B}^{b}{\rm H}^{t}$ (or ${\rm H}^{b}{\rm B}^{t}$) site of the AB-stacking region are not equivalent.
This is evident in Fig.~\ref{fig:EquivalentSites} by examining sites 4, 5 and 6 (depicted by black, gray, and light blue disks, respectively). 
Each of these sites belongs to a different symmetry group within the AB zone due to variations in graphene layer alignment.

\begin{figure}[h!]
    \centering
    \includegraphics[width=0.99\linewidth]{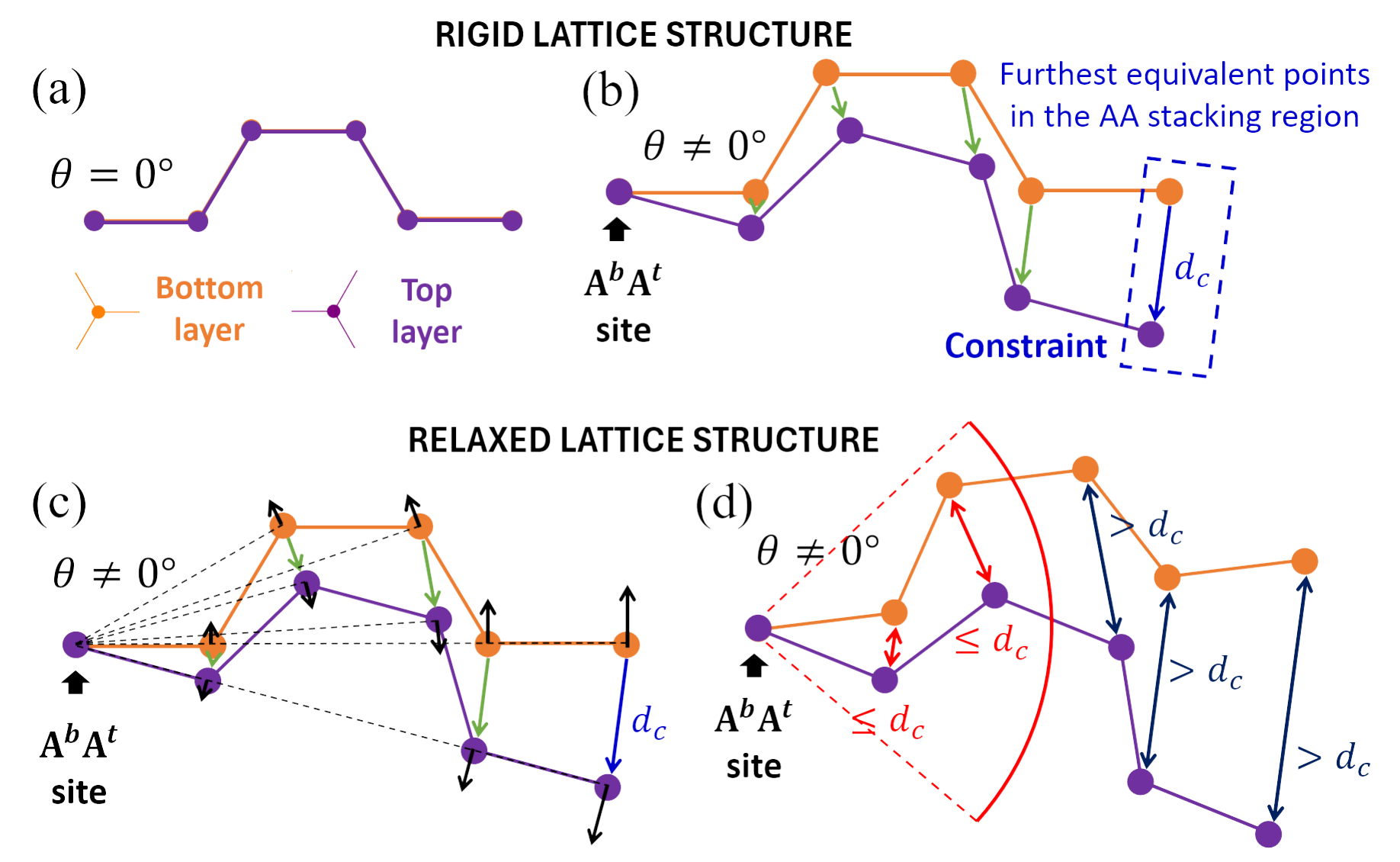}
    \caption{Schematic representation of the lattice structure in the AA-stacking region.
    Purple (orange) circles indicate the carbon sites of the top (bottom) layer.
    The green arrows correspond to the 
    in-plane atomic displacement vectors $\Delta {\mathbf R}_{i}$ produced by lattice relaxation.
    The blue arrow represents the constraint that defines the AA-region area.
    (a) Rigid lattice with $\theta = 0$. 
    (b) Rigid lattice structure with $\theta \neq 0$.
    (c) and (d) Relaxed lattice structures for $\theta \neq 0$.
    The black arrows correspond to displacement vectors due to relaxation.
    The dashed lines indicate the radial distances with respect to site ${\rm A}^{b}{\rm A}^{t}$.
    The dashed and solid red lines correspond to the circular sector of the AA-region area.
    }
    \label{fig:EquivalentAndEquidistantSites}
\end{figure}

Figure \ref{fig:EquivalentAndEquidistantSites} provides a schematic representation of the lattice structure within the AA-stacking region. 
In relaxed lattice structures, we observe a distinctive pattern in the atomic position displacement vectors. Specifically, the $xy$-plane projections of the atomic position displacement vectors, $\Delta {\mathbf R}_{i}(x,y)={\mathbf R}_{i}^{\rm relaxed} - {\mathbf R}_{i}^{\rm rigid}$ [indicated by the black arrows in Fig.~\ref{fig:EquivalentAndEquidistantSites}(c)] are perpendicular to the radial vectors originating from the central ${\rm A}^{b}{\rm A}^{t}$ site.
Moreover, their angular components exhibit opposite directions in the two graphene layers within the AA zone.
Furthermore, the magnitudes of the displacement vectors $\Delta {\mathbf R}_{i}(x,y)$ increase linearly with the radial distance from the central ${\rm A}^{b}{\rm A}^{t}$ site.

The area of the AA-stacking region in a relaxed lattice structure for a given twist angle is evaluated similarly to the rigid lattice case, but with an additional constraint: 
The distance between two furthest equivalent atomic sites belonging to different layers must remain within the upper bound, $d_c$, established for the rigid lattice structure at the same twist angle $\theta$.
This is visually represented by the blue arrow in Figs.~\ref{fig:EquivalentAndEquidistantSites}(b) and (c). 
The atomic displacement distances, represented by green arrows in Fig.~\ref{fig:EquivalentAndEquidistantSites}(c), increase due to lattice relaxation.
Consequently, owing to the constraint imposed by $d_{c}$, the relaxed TBG lattice structures exhibit a smaller AA-stacking region area compared to their rigid counterparts, as illustrated by  Fig.~\ref{fig:EquivalentAndEquidistantSites}(d). 

In summary, we propose 
 {a simple, symmetry-based, and reproducible criterion} 
to determine the AA-stacking area based on the analysis of local symmetry regions. 
This approach offers a clear advantage over the method presented in Ref.~\cite{Zhang2020}, which relies on averaged values obtained from a range of arbitrary cut-off distances.

\subsection{Electronic calculations}
\label{sec:electronic-calculations}

The low-energy electronic properties of  {TBGs} are obtained using a tight-binding model following Nam and Koshino \cite{nam_lattice_2017}:
\begin{equation}
\label{eq:tight-binding}
H=  -\sum_{i,j}  t({\mathbf R}_i - {\mathbf R}_j)  \ket{{\mathbf R}_i}   \bra{ {\mathbf R}_j } + {\rm H. c.} ,
\end{equation}
where ${\mathbf R}_i$ is the position of the $i$th atom and $\bra {\bf r} \ket{{\mathbf R}_i}$ is the corresponding $p_z$ orbital wave function. 
The transfer integral $t({\mathbf R})$ depends on the interatomic distance and the relative orientation between $p_z$ orbitals 
connected by ${\bf R}$.
The transfer integral is parametrized as 
\begin{equation}
-t({\mathbf R}) = V_{pp\pi} (R)  \!\left[ 1-\left( \frac{ {\mathbf R} \cdot  \mathbf{e}_z } {R} \right)^2 \right] +  V_{pp\sigma} (R) \left( \frac{ {\mathbf R} \cdot  \mathbf{e}_z } {R} \right)^2 \!\!,
\end{equation}
where the hopping parameters $V_{pp\pi}(R)$ and $V_{pp\sigma} (R)$ are given by
\begin{equation}
 V_{pp\pi} (R) =  V^0_{pp\pi} \exp \left( -\frac{R - a_0}{ r_0} \right) 
\end{equation}
\begin{equation}
 V_{pp\sigma} (R) =  V^0_{pp\sigma} \exp \left( -\frac{R - d_0}{ r_0} \right)\!,  
\end{equation}
with  $V^0_{pp\pi} = -2.65$~eV and  {$V^0_{pp\sigma} = 0.371$~eV}, 
whereas $r_0= 0.184 \, a$ is the hopping matrix element decay length. 

The electronic local density of states (LDOS), $\rho({\bf r}, \varepsilon)$, of the TBG systems under investigation is computed by direct diagonalization and by the HHK recursive technique \cite{Haydock1980, Vidarte2022, Vidarte2024Q}. 

\subsection{Modeling STM measurements}
\label{sec:STM-theory}

 {
Let us first make explicit what is measured and what we compute. 
In the c-AFM experiments of Ref.~\cite{Zhang2020}, a metallic probe in mechanical contact with the top graphene layer injects a vertical current that flows through the tip/top-layer contact, the interlayer region, and into the bottom electrode. 
Reference~\cite{Zhang2020} models the corresponding conductance semi-empirically as two channels combined in series: 
a Tersoff-Hamann-like term for the tip--surface contact, which carries the local density of states (LDOS) of the top layer, and a purely geometric interlayer term, with empirical parameters originally calibrated for graphene/Ru(0001) contacts~\cite{Song2019}. 
In parallel with these, phonon-assisted processes provide an additional interlayer channel~\cite{Perebeinos2012,Perebeinos2024}. 
}

 {
In what follows, we focus on the tip--surface channel, which carries the dependence of the signal on the evolution of the electronic structure with the twist angle. 
In Appendix~\ref{app:acq}, we evaluate the full two-channel expression of Ref.~\cite{Zhang2020} using our rigid and relaxed geometries, showing that the interlayer term modulates the amplitude but does not control the position of the conductance maximum. 
In Appendix~\ref{app:phonon}, we provide a quantitative bound on the phonon channel.
}

Therefore, we estimate the local conductance using Bardeen's theory \cite{Bardeen1961,Gottlieb2006,Andrei2012} and Tersoff-Hamann formula \cite{Tersoff1983,Tersoff1985,Gottlieb2006},  with a combination of direct diagonalization and the Haydock-Heine-Kelly (HHK) recursion technique \cite{Haydock1980, Vidarte2022, Vidarte2024Q}, considering lattice relaxations obtained from 
semi-empirical forces. 

At low temperatures, the STM current is given by \cite{Bardeen1961,Gottlieb2006,Andrei2012}
\begin{equation}
    I({\bf r}) = \frac{4\pi e}{\hbar} \int_0^{eV} \!\! d\ve \, \rho_{\rm S}({\bf r}, \ve_F - eV {\color{blue}+} \ve) \rho_{\rm T}(\ve_F + \ve) |M|^2 
\end{equation}
where $V$ is the sample-tip bias voltage, $\rho_{\rm S}$ stands for the surface LDOS, $\rho_{\rm T}$ for the tip LDOS, and $M$ is the electron tip-surface tunneling transition matrix element.  
Assuming that the LDOS tip has a very smooth energy dependence, the tunneling current can be approximated as \cite{Andrei2012}
\begin{equation}
\label{Eq:currentM}
    I(\textbf{r}) \propto \left[ \int_0^{eV} \rho_{\rm S}(\textbf{r}, \varepsilon) d\varepsilon \right] e^{-h(\textbf{r})\kappa (\textbf{r})} ,
\end{equation}
where the inverse decay length 
$\kappa \sim \sqrt{2m_e\phi}/\hbar$, with $m_e$ the electron mass, is determined by the local barrier height or average work function $\phi$.
The exponential dependence on tip-sample distance $h({\bf r})$ makes it possible to obtain high-resolution surface topography at a given bias voltage. 
An STM image not only reflects topography but also contains information about the sample LDOS, which can be obtained directly \cite{Andrei2012,Fischer2007} by measuring the local differential conductance
\begin{equation}
    \dfrac{dI({\bf r})}{d{V}} \propto  \rho_{\rm S}({\bf r},\varepsilon = {eV}) ,
\end{equation}
where for simplicity $\ve_{F}$ has been set to zero
in this section (from Sec.~\ref{sec:vertical-current} onwards, $\varepsilon_F$ denotes the Fermi-level position under the tip measured from charge neutrality). 
The calculated local differential conductance, as described above, is a key quantity in this study, since it allows for a direct comparison with STS measurements, as reported in Refs.~\cite{Yu2020, Zhang2020, Liao2018}.

Lattice relaxation modifies $\rho_S({\bf r},\ve_F)$, causing, for instance, the shrinking of the AA-stacking region. 
Out-of-plane relaxation also affects $|M|^2$, which depends exponentially on the tip-surface distance. 

Our STM simulations are conducted in constant height mode. 
We model the LDOS of a given site as a $p_z$ orbital centered on the atomic position, considering only the contribution from the two highest valence bands.
 {The energy range spanned by these two bands depends on the twist angle; the cell-averaged quantities discussed below are instead defined through fixed integration windows, specified in each case.}
As shown below, a 100 meV  {energy window}
effectively captures the salient features of the LDOS within the AA region of the unit cell.
The magnitude of the STM current at a specific point on the surface map is obtained by summing the calculated LDOS contribution from all atomic positions, accounting for the exponential decay of the tunneling current with distance.

 {
A final methodological point concerns the energy scale. 
In contact-mode c-AFM, the energy range of states contributing to the junction conductance is not set by the small applied bias alone, but by the local position of the Fermi level 
under the metallic tip: high-work-function (Pt/Ir-group) contacts dope graphene $p$-type by $0.2$--$0.4$~eV~\cite{Giovannetti2008,Khomyakov2009}, ambient adsorbates provide an additional $p$-doping of $0.1$--$0.3$~eV~\cite{Ryu2010}, and at room temperature the Fermi window is thermally broadened by $\approx 3.5\,k_BT \approx 90$~meV. 
To account for these effects, in Sec.~\ref{sec:vertical-current} we also evaluate the current at the experimental biases and temperature as an explicit function of the Fermi-level position $\varepsilon_F$,
\begin{equation}
\label{Eq:finiteT-current}
\begin{split}
G(\theta;\varepsilon_F,V,T) \propto{}& \frac{1}{eV}\int d\varepsilon\; \bar{\rho}_{S}(\varepsilon) \\
&\times \left[f(\varepsilon-\varepsilon_F-eV)-f(\varepsilon-\varepsilon_F)\right],
\end{split}
\end{equation}
where $\bar{\rho}_{S}$ is the cell-averaged surface LDOS and $f$ is the Fermi function. 
The fixed integration windows used below correspond to the zero-temperature limit of Eq.~\eqref{Eq:finiteT-current} at charge neutrality.}

\section{Results}
\label{sec:results}

\subsection{Structural relaxation}
\label{sec:structural-relaxation}

We will now focus on examining the out- and in-plane relaxation of TBG lattice structures, obtained by the LAMMPS simulations. 
Here, we consider the $\vert n-m\vert =1$ family of odd commensurate lattice structures.

\subsubsection{Out-of-plane relaxation}
\label{sec:out-of-plane}

\begin{figure}[h!]
    \centering
    \includegraphics[width=1\linewidth]{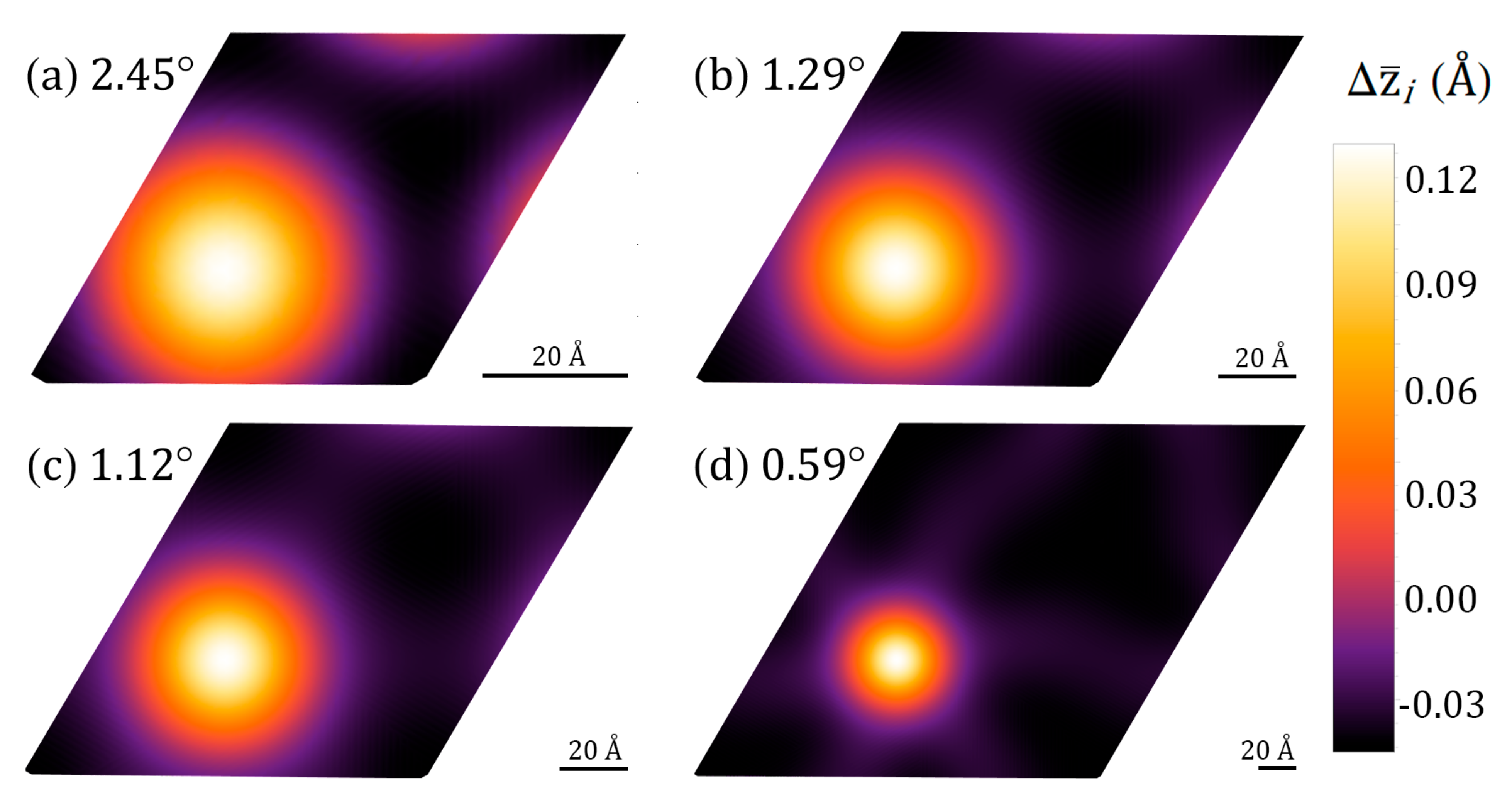}
    \caption{
    Spatial representation of the average out-of-plane displacement $\Delta \bar{z}_{i}$ within the moir\'e unit cell for the twist angles (a) $\theta\approx 2.45^{\circ}$, (b) $1.29^{\circ}$, (c) $1.12^{\circ}$, and (d) $0.60^{\circ}$.
    The color scale bar indicates the values of $\Delta \bar{z}_{i}$ (in angstrom). 
    }
    \label{fig:delta z 2D plot}
\end{figure}

\begin{figure}[h!]
    \centering
    \includegraphics[width=0.65\linewidth]{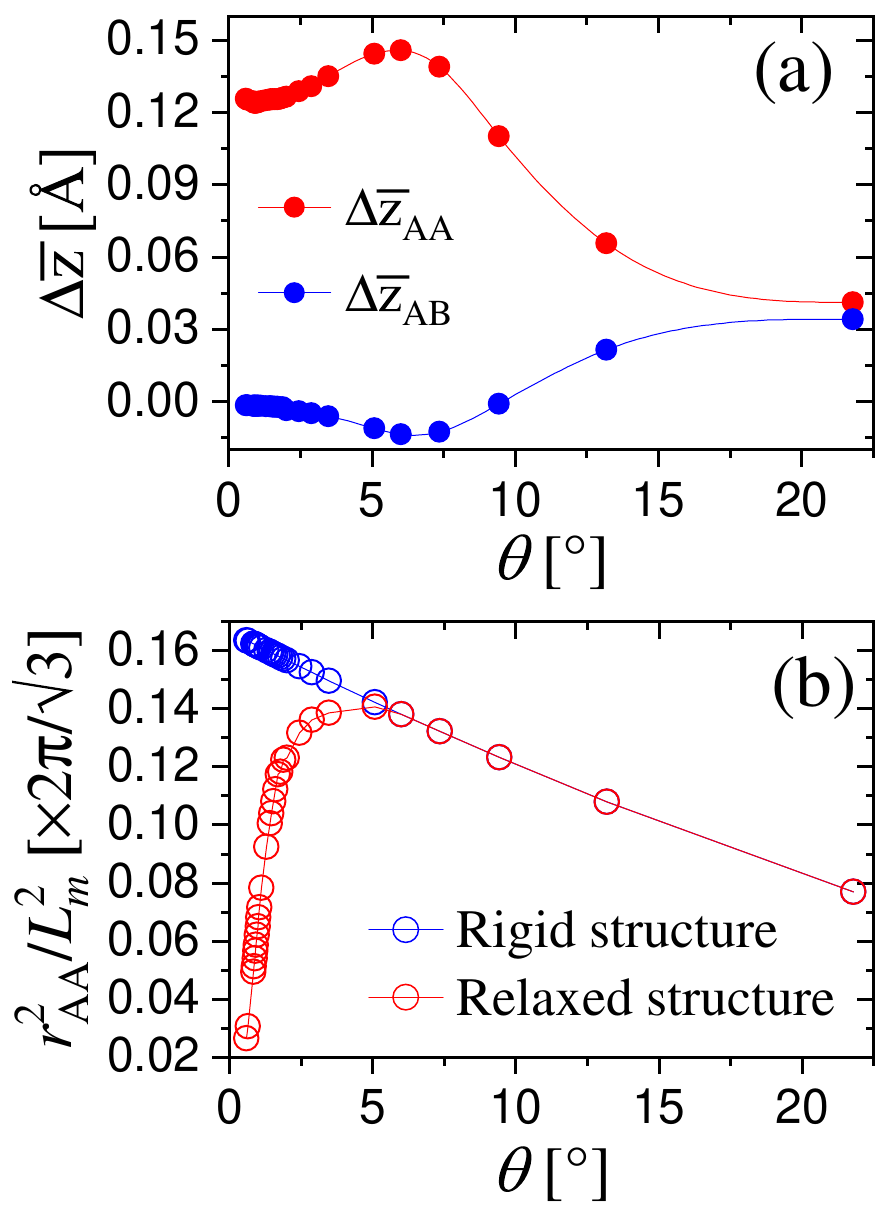}
    \caption{
    (a) Out-of-plane displacements $\Delta \bar{z}_{\rm AA}$ and $\Delta \bar{z}_{\rm AB}$ (in angstrom) at the ${\rm A}^{b}{\rm A}^{t}$ and ${\rm H}^{b}{\rm B}^{t}$ atomic sites, respectively, of the top layer as a function of twist angle $\theta$ (in degrees).
    (b) The ratio between the AA-region area and the area of the moir\'e unit cell, $\propto r_{\rm AA}^{2}/L_{m}^{2}$, as a function of $\theta$ (in degrees).
    The blue and red open points correspond to the rigid and relaxed lattice structure, respectively.
    }
    \label{fig:Relaxed}
\end{figure}

We quantify the out-of-plane relaxation by the average vertical displacement of each top-layer site with respect to its unrelaxed plane, defined as
\begin{equation}
    \Delta \bar{z}_{i}=\frac{1}{2N_b}\sum_{j=1}^{N_b} \left(z_{i,{\rm top}}^{\rm relaxed} - z_{j,{\rm bottom}}^{\rm relaxed} - d_{0}\right),
    \label{eq:deltaz}
\end{equation}
where, for each atomic site $i$ of the top layer, the average runs over the $N_b$ atoms $j$ of the bottom layer within a circular radius of $5$~\AA~centered on $i$.
The two layers relax symmetrically about the midplane of the bilayer, to a very good approximation, so that $\Delta \bar{z}_{i}$ is the vertical displacement of the top-layer atom (the quantity shown in Figs.~\ref{fig:delta z 2D plot} and \ref{fig:Relaxed}), whereas the local change of the interlayer separation with respect to $d_0$ is $2\Delta \bar{z}_{i}$.

Figure~\ref{fig:delta z 2D plot} presents contour plots illustrating the out-of-plane average 
displacement $\Delta \bar{z}_{i}$ within the moir\'e unit cell of TBGs with $\theta\approx 2.45^{\circ}$, $1.29^{\circ}$, $1.12^{\circ}$ and $0.60^{\circ}$, encompassing 2188, 7804, 10444 and 38308 carbon atoms, respectively.
The bright areas ($\Delta \bar{z}_{i}>0$) correspond to atomic sites with the largest out-of-plane displacements following lattice structural relaxation. 
These areas coincide with the AA-stacking regions, specifically at one-third of the largest diagonal of the unit cell.
Conversely, the dark areas ($\Delta \bar{z}_{i}<0$) correspond to the AB-stacking regions located at the vertices of the unit cell and at two-thirds of its larger diagonal.

To highlight the atomic sites with the most prominent enhancement of out-of-plane displacement in TBGs, we consider the ${\rm A}^{b}{\rm A}^{t}$ and ${\rm H}^{b}{\rm B}^{t}$ (or ${\rm B}^{b}{\rm H}^{t}$) sites as representative sites of the AA- and AB-stacking regions, respectively.
Figures~\ref{fig:Relaxed}(a) and (b) illustrate the evolution of the out-of-plane displacement, twist angle $\theta$ 
at the ${\rm A}^{b}{\rm A}^{t}$ and ${\rm H}^{b}{\rm B}^{t}$ atomic sites, respectively.



As shown in Fig.~\ref{fig:Relaxed}(a), decreasing the twist angle increases the out-of-plane displacement of the AA-stacking region ($\Delta\bar{z}_{\rm AA}>0$), which saturates for $\theta\lesssim 5^{\circ}$ at $\sim 0.15$~\AA, displaying a shallow maximum near $\theta\approx 5^{\circ}$ [$(m,n)=(6,7)$, $508$ carbon atoms in the moir\'e cell].
Since both layers bulge outwards by this amount, at this saturation value the interlayer separation in the AA-stacking region exceeds the spacing $d_0$ of the non-relaxed lattice by $\approx 2\times 0.15 = 0.30$~\AA, in agreement with previous relaxation studies~\cite{nam_lattice_2017}.

{
Indeed, the local interlayer distances measured on our relaxed structures
 are $d_{\rm AA} = 3.66\text{ \AA}$ ($3.62\text{ \AA}$) and $d_{\rm AB} = 3.35\text{ \AA}$ ($3.37\text{ \AA}$) at $\theta = 5.09^{\circ}$ ($1.12^{\circ}$), in very good agreement
with published full-relaxation results~\cite{Uchida2014,nam_lattice_2017}.
}
The shallow maximum reflects the competition between the larger moir\'e cell---which provides more room for the AA core to buckle out---and the simultaneous shrinking of the AA-stacking area as $\theta$ decreases.
In contrast, the AB-stacking regions exhibit an interlayer spacing that exceeds $d_{0}$ ($\Delta\bar{z}_{\rm AB}>0$) for $\theta >10^{\circ}$.
However, for $\theta <10^{\circ}$ the graphene layers in the AB-stacking are brought closer together, with a spacing smaller than $d_{0}$ ($\Delta\bar{z}_{\rm AB}<0$).
The minimum displacement of $\sim 0.014$~\AA~occurs close to $\theta= 5^{\circ}$, indicating that the AB-stacking regions are closer by $\sim 0.028$~\AA .

\subsubsection{In-plane relaxation}
\label{sec:in-plane}

We can evaluate the horizontal stress of the lattice by calculating the differences between the atomic positions in the rigid and relaxed structures. 
The in-plane displacement vectors $\Delta {\mathbf R}_{i}(x,y)={\mathbf R}_{i}^{\rm relaxed} - {\mathbf R}_{i}^{\rm rigid}$ reach a maximum at the boundary of the AA-stacking region, which assumes a circular shape.
Our calculations indicate that the carbon-carbon nearest neighbor bond modifications are negligible for $\theta \agt 5^{\circ}$.
However, as the twist angle decreases below $\theta \alt 5^{\circ}$, the carbon-carbon nearest neighbor bond modifications increase significantly. Carbon atoms are increasingly displaced to restore AB- and BA-stacking, which assumes triangular domains.

Let us now examine the AA-region area of the rigid and relaxed lattice structures using the procedure described in Sec.~\ref{sec:AA-region-area} based on point group symmetries. Figure~\ref{fig:Relaxed}(b) presents the ratio between the AA-region area, proportional to $ r_{\rm AA}^{2}$, and the area of the moir\'e cell,  proportional to $L_{m}^{2}$, as a function of twist angle $\theta$ for both rigid and relaxed lattice structures.
Given the $1/\theta^{2}$ dependence of the moiré cell area at small twist angles, see Eq.~\eqref{Eq:lattice_constant}, the ratio $r_{\rm AA}^{2}/L_{m}^{2}$ exhibits a linear dependence for rigid lattice structures.
However, the calculations show a clear deviation from this linear behavior for $\theta <5^{\circ}$ upon lattice relaxation.
This deviation indicates a significant reduction in the AA-region area within moiré unit cells with $\theta <5^{\circ}$, as further corroborated by Fig.~\ref{fig:delta z 2D plot}. Consequently, the triangular domains of the AB- and BA-stacking regions expand.

In summary, in the absence of lattice relaxation, a decrease in twist angle inherently leads to an increase in AA-stacking region size.
However, given that AA stacking is energetically less favorable than AB stacking, the system undergoes lattice relaxation to minimize these high-energy areas.
This minimization involves in-plane relaxation, which contracts AA regions by pulling edge atoms towards the more stable AB configuration, consequently expanding the AB and BA domains. 
Simultaneously, out-of-plane relaxation causes an increased interlayer separation within AA regions, thereby relieving the associated strain. 
The 5$^{\circ}$ twist angle represents a critical point where the system balances the competing effects of minimizing AA area and accommodating the associated strain through a combination of  out-of-plane and in-plane relaxations.
This balance is reflected in the maximum interlayer separation observed in AA regions and the minimum separation in AB regions around this angle.


This represents the prevailing and appealing explanation for the contact conductance peak at a $\theta = 5^\circ$ twist angle. However, through a careful analysis of the LDOS and detailed modeling of STM measurements, we demonstrate in the following that this is not the dominant mechanism responsible for the observed peak.
 {The distinct contributions of the in-plane and out-of-plane structural relaxations to the LDOS and to the contact current are quantified in Appendix~\ref{app:ipop}.}

\subsection{Analysis of STM in constant height mode}
\label{sec:STM}

We now apply the theoretical framework and methods outlined in Sections~\ref{sec:method_and_methods} to analyze the vertical current reported in the STM experiment \cite{Zhang2020}.

\begin{figure}[h!]
    \centering
    \includegraphics[width=1\linewidth]{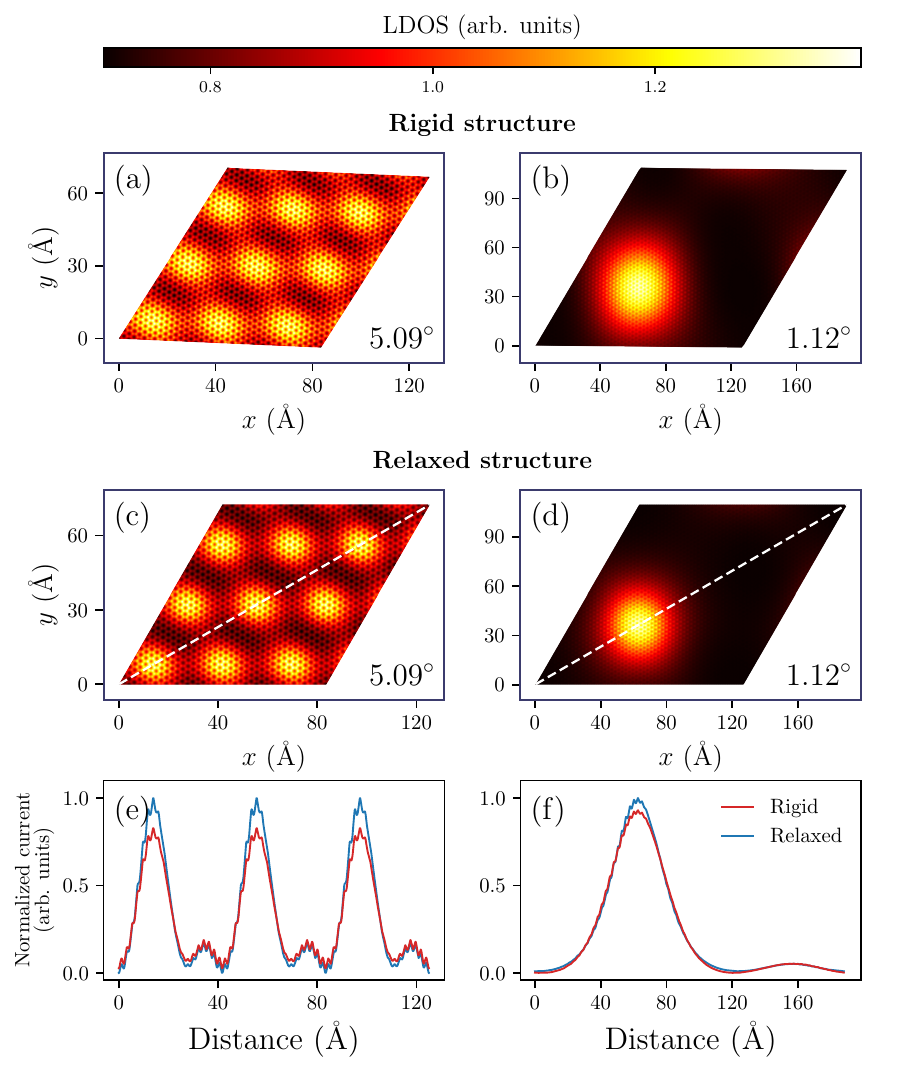}
    \caption{STM simulations in constant height mode for $\theta\approx 5.09^{\circ}$ and $1.12^{\circ}$.
    Panels (a) and (b) show LDOS (in arbitrary units) for rigid structures, while (c) and (d) for relaxed ones.  
    The white dashed lines in panels (c) and (d) indicate the paths along which the normalized current profiles are calculated. 
    Panels (e) and (f) show the normalized current (in arbitrary units) along the path for the relaxed structures at $\theta\approx 5.09^{\circ}$ and $1.12^{\circ}$, respectively. 
    The simulations consider the states belonging to the two highest valence bands for each structure.}
    \label{fig:STM simulations}
\end{figure}

We begin by examining the impact of in-plane relaxation on contact conductance.
Figures~\ref{fig:STM simulations}(a) to (d) present mappings of the LDOS obtained by STM simulations with constant height mode at $1$~\AA~ from the surface, based on the tight-binding model given by Eq.~\eqref{eq:tight-binding}. 
These simulations are performed for twisted angles of $5.09^\circ$ and $1.12^\circ$, considering both rigid [panels (a) and (b)] and relaxed crystalline structures [panels (c) and (d)]. 
For $\theta\approx 5.09^\circ$, $3\times3$ superstructures are plotted in Fig.~\ref{fig:STM simulations}(a) and (c). 
The shrinking of the AA-stacking regions [bright zones in Fig.~\ref{fig:STM simulations} (a)-(d)] is evident when comparing both relaxed and rigid configurations across all panels, and the STM simulations are consistent with the experimental findings \cite{Zhang2020}.
This reduction underscores the significant impact of in-plane lattice relaxation, which tends to suppress AA-stacking areas due to its inherent instability, as discussed in Sec.~\ref{sec:structural-calculations}.

Furthermore, we simulate the current (or contact conductance) profile as the tip moves along the main diagonal of the unit cell. 
Figures~\ref{fig:STM simulations}(e) and (f) display the normalized current measured along the main diagonal [white dashed lines in Fig.~\ref{fig:STM simulations}(c) and (d)] of the unit cell for $\theta\approx 5.09^{\circ}$ and $1.12^{\circ}$, respectively.
The reported currents are normalized to a maximum value of 1 in arbitrary units. 
The highest normalized current value is obtained around the AA-stacking region, consistent with the enhanced LDOS in these areas, which amplifies the vertical current. 
Conversely, regions with low normalized current correspond to AB- and BA-stacking regions.

We highlight two important aspects concerning the current in the AA region.  
Firstly, the current for the relaxed structure at the AA region is roughly 10\% larger than that for the rigid one, as illustrated by Figs.~\ref{fig:STM simulations}(e) and (f).
This can be understood by considering that relaxation compresses the AA region, and thus the states within the energy range between the Fermi level and the tip bias value become confined to a reduced area, increasing the current. 
As previously noted, the current calculation includes contributions from all atoms on the surface. 
More importantly, given that the current is calculated in constant height mode, and considering our prior finding that the AA zone slightly protrudes out of the plane in the relaxed structure, this out-of-plane displacement further enhances the observed current.
 {In addition, one might intuitively expect the peak to broaden for the rigid structure, however, no such broadening is observed.
}
We attribute this to our current calculation methodology, which considers the contribution of all surface atoms, not merely those directly beneath the tip. 

It is important to note that only surface atoms are explicitly considered in the current calculation, since the contribution from the lower layer is deemed negligible due to the exponential decay of electron wave functions. 
The influence of the lower layer is implicitly accounted for in the wave function of the surface atoms, similar to observations in graphite or bilayer graphene with a triangular lattice \cite{stm}.



\subsection{Vertical transport, contact conductance and local density of states.}
\label{sec:vertical-current}

We now analyze the maximum in the contact conductance observed near $5^{\circ}$ and elucidate its origin.
For each twist angle $\theta$, and for both rigid and relaxed structures, we evaluate the tunneling current at $1$~\AA~above the top layer and average it over the moir\'e unit cell.
 {
This cell-averaged current captures the angular dependence of the dominant tip/top-layer channel of the junction probed in the c-AFM experiments~\cite{Zhang2020} (the full two-channel expression is evaluated in Appendix~\ref{app:acq})
}, and it is proportional to the surface LDOS integrated from 
$\varepsilon_F - eV_{\rm bias}$ 
up to the Fermi level [the integral within the bracket of Eq.~\eqref{Eq:currentM}].

\begin{figure}[h!]
\centering
\includegraphics[width=1.0\columnwidth]{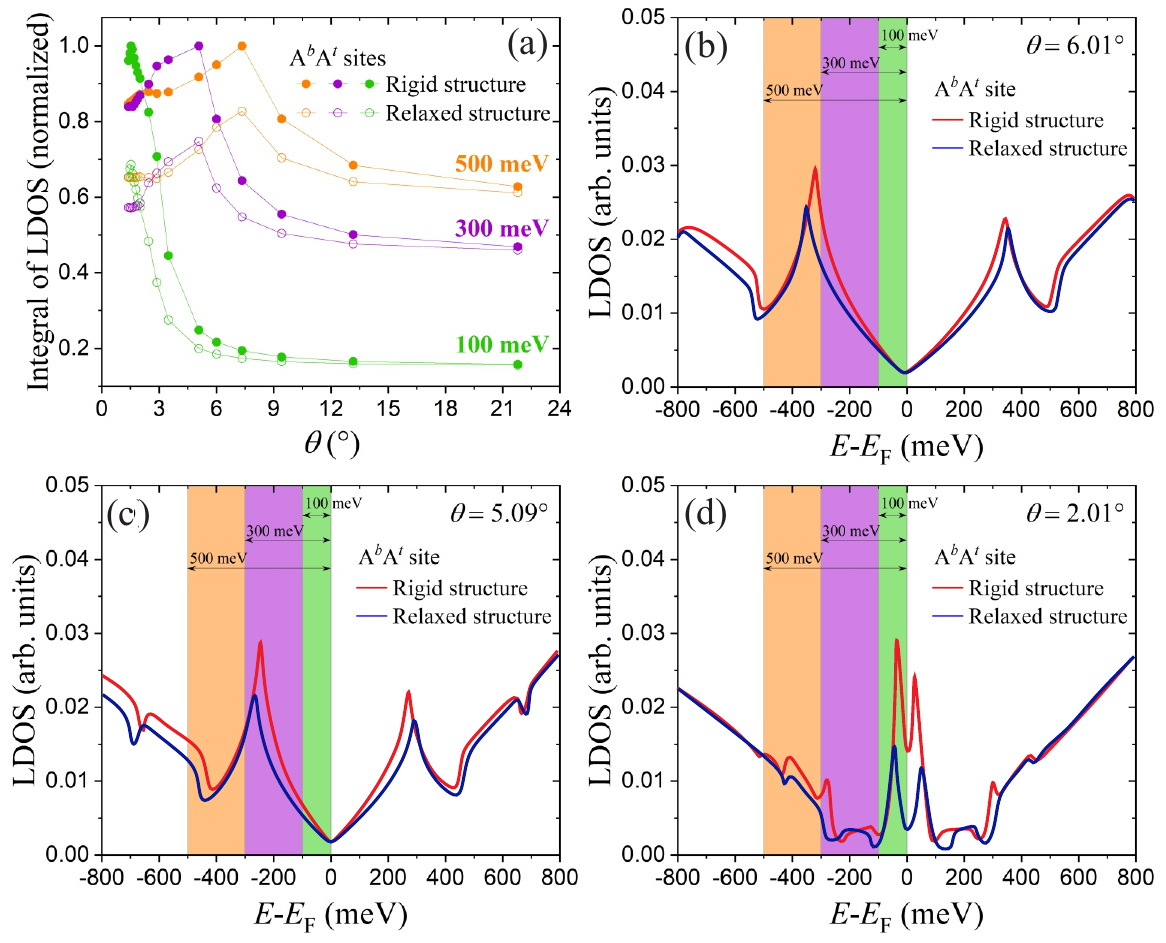}
\caption{(a) Cell-averaged surface current (normalized) at $1$~\AA{} above the top layer as a function of the twist angle $\theta$, obtained by integrating the surface LDOS over energy windows of $100$, $300$, and $500$~meV below the Fermi level (green, purple, and orange, respectively). Filled (open) symbols denote rigid (relaxed) structures. The maximum is non-monotonic and shifts with the energy window, while persisting in the rigid structure.
(b)--(d) LDOS at the ${\rm A}^{b}{\rm A}^{t}$ site for $\theta=6.01^{\circ}$, $5.09^{\circ}$, and $2.01^{\circ}$, respectively, for rigid (red) and relaxed (blue) structures. The shaded regions mark the $100$, $300$, and $500$~meV integration windows used in panel (a).
}
\label{fig:LDC}
\end{figure}

Figure~\ref{fig:LDC}(a) shows the cell-averaged current as a function of the twist angle $\theta$ for rigid and relaxed structures, for three integration windows 
$eV_{\rm bias}=100$, $300$, and $500$~meV below the Fermi level.
In contrast to the monotonic trend expected from interlayer coupling alone, the averaged current is clearly non-monotonic: it grows as $\theta$ decreases, reaches a maximum, and then drops toward the magic angle, reproducing the anomalous behavior measured in Ref.~\cite{Zhang2020}.
The position of this maximum shifts systematically with the energy window: $\theta \approx 7^{\circ}$ for $eV_{\rm bias}=500$~meV, $\theta \approx 5^{\circ}$ for $eV_{\rm bias}=300$~meV, and $\theta \approx 2.5^{\circ}$ for $eV_{\rm bias}=100$~meV.
Crucially, the maximum is present for both rigid and relaxed structures and occurs at essentially the same angle---relaxation shifts it by less than $1^{\circ}$---which already indicates that the peak is not created by lattice relaxation.

 {We emphasize that the geometric AA-area fraction and the energy-integrated surface LDOS are logically independent quantities, and neither should be treated as interchangeable evidence for or against reconstruction: within the tunneling description, the current is controlled by the spectrum of surface states participating in the contact, not by the geometric area of any stacking region. }

 {
In the rigid lattice the AA-area fraction is strictly monotonic in $\theta$, yet the van Hove singularities still shift with the twist angle, yielding a non-monotonic integrated surface LDOS. 
This is precisely why the rigid lattice provides the discriminating benchmark: the persistence of the maximum in the rigid case proves that the peak is spectral (electronic) in origin.}

 {
We stress that the relevant energy scale is not the small applied bias but the range of surface states that contribute to the contact conductance, which is governed by the local tip-induced doping (Sec.~\ref{sec:STM-theory}). 
This point is made quantitative below, where the current is evaluated via Eq.~\eqref{Eq:finiteT-current} at the experimental biases and temperature as a function of the Fermi-level position.
}

To understand this behavior, we present in Figs.~\ref{fig:LDC}(b)-(d) the LDOS at the AA site for three representative angles, for both rigid and relaxed structures.
The highlighted regions indicate the integration windows.
Although the LDOS of the two structure types differs only slightly, the position of the averaged-current maximum tracks the energy of the van Hove singularities (vHs), which have been directly resolved by STM and shift toward the Fermi level as $\theta$ decreases~\cite{Brihuega2012}, and whether they fall within the window between $-V_{\rm bias}$ and the Fermi level.
For $\theta\approx6^{\circ}$ the vHs enters the window only for $V_{\rm bias}=500$~meV, whereas for $V_{\rm bias}=300$~meV the vHs enters the window only once $\theta\lesssim5^{\circ}$, accounting for the bias-dependent shift of the maximum.

This bias dependence is the key electronic signature: a purely geometric mechanism tied to the AA-stacking area would pin the maximum to a single angle, independent of $V_{\rm bias}$.
The fact that the maximum is reproduced---at the same angle---by the rigid structure further confirms that it originates from the evolution of the electronic structure, namely the vHs approaching the Fermi level as $\theta$ decreases, rather than from lattice relaxation or the growth of the AA-stacking area~\cite{Zhang2020}.
Relaxation modulates the effect but does not create it.

\begin{figure*}[t!]
\centering
\includegraphics[width=0.85\textwidth]{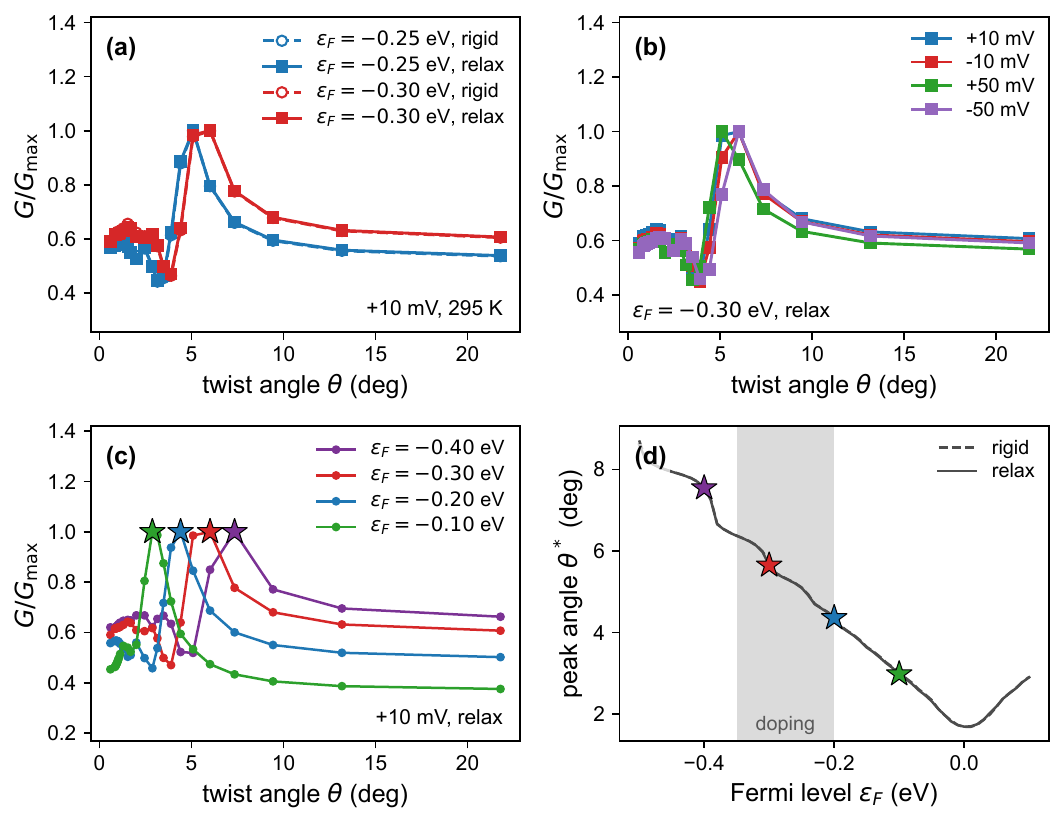}
\caption{
 {Contact current evaluated under experimental conditions, Eq.~\eqref{Eq:finiteT-current}, at $T=295$~K.
(a) Normalized conductance $G(\theta)$ at $V=+10$~mV for $\varepsilon_F=-0.25$ and $-0.30$~eV; open (filled) symbols denote rigid (relaxed) structures, which are indistinguishable on this scale.
(b) Polarity and bias robustness: $G(\theta)$ at $\varepsilon_F=-0.30$~eV for $V = \pm10\text{ mV}$ and $\pm50\text{ mV}$.
(c) Construction of the peak-angle map: normalized $G(\theta)$ at four representative Fermi-level positions ($V=+10$~mV, relaxed structures); stars mark the maximum $\theta^{*}$ of each curve.
(d) Angle of maximum conductance $\theta^{*}$ versus the Fermi-level position (solid line: relaxed; dashed line: rigid, curves are visually indistinguishable). 
Each star corresponds to the maximum of the respective curve in (c). 
The shaded band marks the doping range $-0.35$ to $-0.20$~eV consistent with the contact-electrostatics estimate for Pt-group tip coatings (for details, see Appendix~\ref{app:doping}).}}
\label{fig:finiteT}
\end{figure*}

{Figure~\ref{fig:finiteT} shows the current evaluated at the experimentally relevant biases and temperature, Eq.~\eqref{Eq:finiteT-current}, as an explicit function of the Fermi-level position under the tip.
For any $\varepsilon_F$ within the range $-0.35~{\rm eV} \le \varepsilon_F \le -0.20$~eV expected for Pt-group tip coatings (see Appendix~\ref{app:doping} for details), the conductance maximum lies at $\theta^{*}=4.4^{\circ}$--$6.5^{\circ}$ for $V=\pm10$~mV, with the two polarities agreeing to better than $0.4^{\circ}$; the $\pm50$~mV results follow the same trend [Fig.~\ref{fig:finiteT}(b)].
Remarkably, the rigid and relaxed structures yield the same $\theta^{*}$ to better than $0.06^{\circ}$ 
at every doping level.
No fine-tuning is involved: any Fermi-level position within 
this range places the maximum within the $4.4^{\circ}$--$6.5^{\circ}$ interval}, 
and $\theta^{*}(\varepsilon_F)$ defines a smooth, monotonic map [Figs.~\ref{fig:finiteT}(c) and \ref{fig:finiteT}(d)].
This systematic evolution provides a clear fingerprint of the electronic mechanism: under electrostatic gating or controlled doping, the conductance maximum is predicted to shift in angle, whereas a mechanism originating purely from the AA-stacking area would pin it.

The logical structure of this argument deserves emphasis, since it clearly distinguishes what is established parameter-free from what depends on the doping level.

The existence of the conductance maximum, its electronic origin, and the inadequacy of a purely structural explanation follow from the comparison between rigid and relaxed structures and from the analyses of Appendices~\ref{app:acq}, \ref{app:ipop} and \ref{app:phonon}, none of which involve any adjustable parameters.

The exact position of the maximum depends on a single physical quantity, the local Fermi-level position under the tip, which was not measured in Ref.~\cite{Zhang2020} and which we do not fit.
Its range is the one expected for a high-work-function metal in contact with graphene~\cite{Giovannetti2008,Khomyakov2009}, consistent with direct measurements of the graphene work function and of the Fermi-level shifts induced by metal contacts and by scanning metallic probes~\cite{Yu2009,Panchal2013,Song2012,Pletikosic2009,Choudhary2011,Zhao2015}. 
Our self-consistent estimate of the contact-induced Fermi-level shift for the tip/TBG geometry, obtained by solving the contact electrostatics of Refs.~\cite{Giovannetti2008,Khomyakov2009} with the density of states of the twisted bilayer itself, shows that the shift is essentially independent of the twist angle (it varies by less than $70$~meV between $0.6^{\circ}$ and $21.8^{\circ}$) and of relaxation (rigid and relaxed structures agree to within $8$~meV). 
This result fully justifies the use of a single $\varepsilon_F$ for all angles.
The magnitude of $\varepsilon_F$ is controlled by the work function $W_M$ of the tip coating. 
The bare offset $W_M-W_G-\Delta_c$, with $W_G=4.48$~eV the graphene work function and $\Delta_c\approx0.9$~eV the interface-dipole step of Ref.~\cite{Khomyakov2009}, changes sign at $W_M\approx5.4$~eV and is partly screened by the transferred charge. 
For a metal-graphene separation of $3.3\text{ \AA}$, our self-consistent model yields $\varepsilon_F\approx-0.10$~eV for polycrystalline Ir ($W_M=5.5$~eV), $-0.26$~eV for $W_M=5.8$~eV, and $-0.33$~eV for $W_M=6.0$~eV [Fig.~\ref{fig:doping}(b)], in agreement with the reported work functions of Pt ranging from $5.65$~eV (polycrystalline) to $6.1$~eV [Pt(111)]~\cite{Michaelson1977,Derry1989}.
Consequently, the doping range that places the calculated maximum within the observed angular window, $-0.30\lesssim\varepsilon_F\lesssim-0.20$~eV [Fig.~\ref{fig:finiteT}(d)], is thus reached for Pt-rich coatings and for the upper part of the Ir range, whereas the lowest reported Ir values would leave the bilayer nearly undoped (Appendix~\ref{app:doping}).
Finally, we emphasize that the same value of $\varepsilon_F$ must simultaneously account for three distinct features: the position of the maximum, the angular contrast, defined as the ratio $G(\theta^{*})/G(21.8^{\circ})$ between the conductance at the maximum and at the largest angle considered ($1.6$--$2.0$ calculated over the range of Fig.~\ref{fig:finiteT}(d); $1.5$--$1.9$ measured in Fig.~2 of Ref.~\cite{Zhang2020}, where the current is normalized to the domain with $\theta>12^{\circ}$), and the conductance decay towards small angles.
The theoretical description is therefore highly over-constrained rather than fitted.

It is worth noting that the carrier density at ${\rm H}^{b}{\rm B}^{t}$ site as a function of the angle, exhibits a similar trend to that at ${\rm A}^{b}{\rm A}^{t}$ site, but with reduced intensity.
All results presented in this section have been obtained using both direct diagonalization and the HHK recursive technique \cite{Vidarte2022,Vidarte2024Q,Vidarte2025}, considering both relaxed and rigid lattice structures.

\section{Conclusions and discussion}
\label{sec:conclusion}

In this study, we investigate the intriguing behavior of contact conductance in TBG systems, providing a  {theoretical framework that addresses} recent, unexpected experimental observations from STM and c-AFM.
These experiments highlighted a surprising non-monotonic current pattern dependent on the TBG rotation angle 
$\theta$, with a prominent peak around $\theta =5 ^\circ$. 
This finding presented a significant departure from the widely recognized magic angle behavior of TBG, prompting our comprehensive investigation.

Our theoretical and computational approach, applied to both relaxed and rigid TBG structures, allowed us to simulate contact conductance by systematically analyzing the LDOS across various biases and rotational angles. 
A key insight from our work is a reassessment of the prevailing interpretation that links the observed conductance maximum to structural relaxation or changes in the AA-stacking area.
While structural relaxation indeed plays a role in modifying the physical dimensions of stacking regions, our simulations demonstrate that it is not the primary driver of the conductance peak near $\theta = 5^\circ$: the cell-averaged contact current exhibits the same non-monotonic maximum in the rigid (unreconstructed) structure, with relaxation shifting its position by less than $1^\circ$.
Instead, our findings  {indicate} that the conductance maximum originates from the evolution of the electronic band structure.
Specifically, we identified the shifting of van Hove singularities (vHs) toward the Fermi level as the underlying mechanism as the twist angle decreases. This energetic alignment of vHs with the Fermi level has a profound impact on the available electronic states for tunneling, leading to the observed conductance enhancement.

 {Furthermore, the precise location of the conductance maximum depends on the energy window of states contributing to the contact, which is governed by the local tip-induced doping. 
This interplay between twist angle, doping, and the energy of the vHs accounts for the experimental observations and leads to a falsifiable prediction: under electrostatic gating or controlled doping, the conductance maximum should shift in angle, whereas a purely structural mechanism would pin it.}
 {Our results clarify the previously puzzling experimental data and reveal the dominant role of the electronic band structure on the vertical transport properties of TBG.}
This understanding is crucial for designing and optimizing future graphene-based electronic devices, particularly those that rely on tunable contact properties.

\appendix

\section{Full two-channel contact conductance}
\label{app:acq}

In this Appendix we evaluate the complete semi-empirical contact-conductance expression of Ref.~\cite{Zhang2020}, in which the tip--surface and interlayer channels are combined in series (inverse-sum form) for every (top) surface atom $j$,
\begin{equation}
\label{Eq:acq}
G = \Big\langle \sum_{j\in{\rm top}} \Big[ \big(G^{\rm tip}_{j}\big)^{-1} + \big(G^{\rm int}_{j}\big)^{-1} \Big]^{-1} \Big\rangle_{\rm cell},
\end{equation}
with
\begin{equation}
\label{Eq:acq-channels}
G^{\rm tip}_{j} = \rho_{j}\, e^{-3.91\, d_{j\text{-tip}} + 10},
\quad
G^{\rm int}_{j} = \sum_{i\in{\rm bot}} e^{-4.20\, d_{ji} + 10.93},
\end{equation}
where the empirical constants are those of Refs.~\cite{Zhang2020,Song2019} (originally calibrated for graphene/Ru(0001) contacts), 
$d_{j\text{-tip}}$ and $d_{ji}$ are the distances, in \AA, from atom $j$ of the top layer to the tip apex and to atom $i$ of the bottom layer (the exponents are in \AA$^{-1}$, so that $G$ is in the arbitrary units of Ref.~\cite{Zhang2020}),
$\rho_j$ is the per-atom LDOS integrated over the same energy windows as in Fig.~\ref{fig:LDC}, the tip is placed $1$~\AA{} above the local top layer, and periodic images are included up to $10$~\AA.

\begin{figure*}[t!]
\centering
\includegraphics[width=0.9\textwidth]{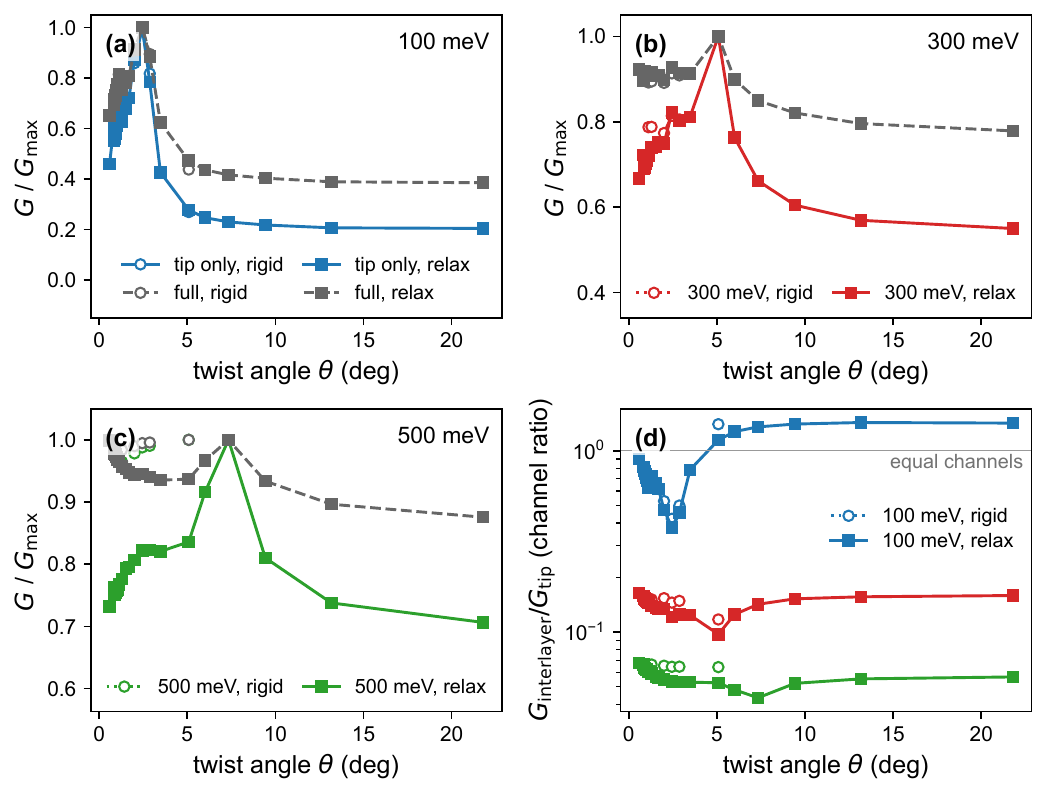}
\caption{
Cell-averaged contact conductance versus twist angle $\theta$ computed with the tip--surface term alone (color) and with the full two-channel expression of Eq.~\eqref{Eq:acq} (gray), for rigid (open symbols, dashed lines) and relaxed (filled symbols, solid lines) structures and integration windows of (a) $100$, (b) $300$, and (c) $500$~meV. 
(d) Ratio of the interlayer to the tip--surface channel; the interlayer channel is $\theta$-independent for the rigid lattice and smoothly monotonic for the relaxed lattice. 
The $300$ and $500$~meV pairs in panel (d) follow the line styles identified in the legends of panels (b) and (c).
}
\label{fig:acq}
\end{figure*}

Figure~\ref{fig:acq} compares, for rigid and relaxed structures and for the three windows, the cell-averaged conductance computed with the tip--surface term alone and with the full two-channel expression.
The position of the maximum is unchanged by the inclusion of the interlayer term ($\theta^{*}\approx 2.5^{\circ}$, $5.1^{\circ}$, and $7.3^{\circ}$ for the $100$, $300$, and $500$~meV windows, respectively, in all cases, within our angular sampling of $\lesssim 0.7^{\circ}$); the interlayer term only compresses the amplitude contrast.
The reason is purely geometrical: for the rigid (unreconstructed) lattice the interlayer channel per unit area is independent of the twist angle to within $5\times10^{-6}\,\%$ over the full range $0.99^{\circ}$--$21.8^{\circ}$, whereas for the relaxed structures it varies smoothly and monotonically by $\approx 19\%$, reflecting the increased AA interlayer distance.
A channel that is constant or smoothly monotonic in $\theta$ cannot generate a conductance maximum; combined in series, it merely modulates the amplitude of the maximum produced by the tip--surface channel without displacing it.

\begin{figure*}[t!]
\centering
\includegraphics[width=0.85\textwidth]{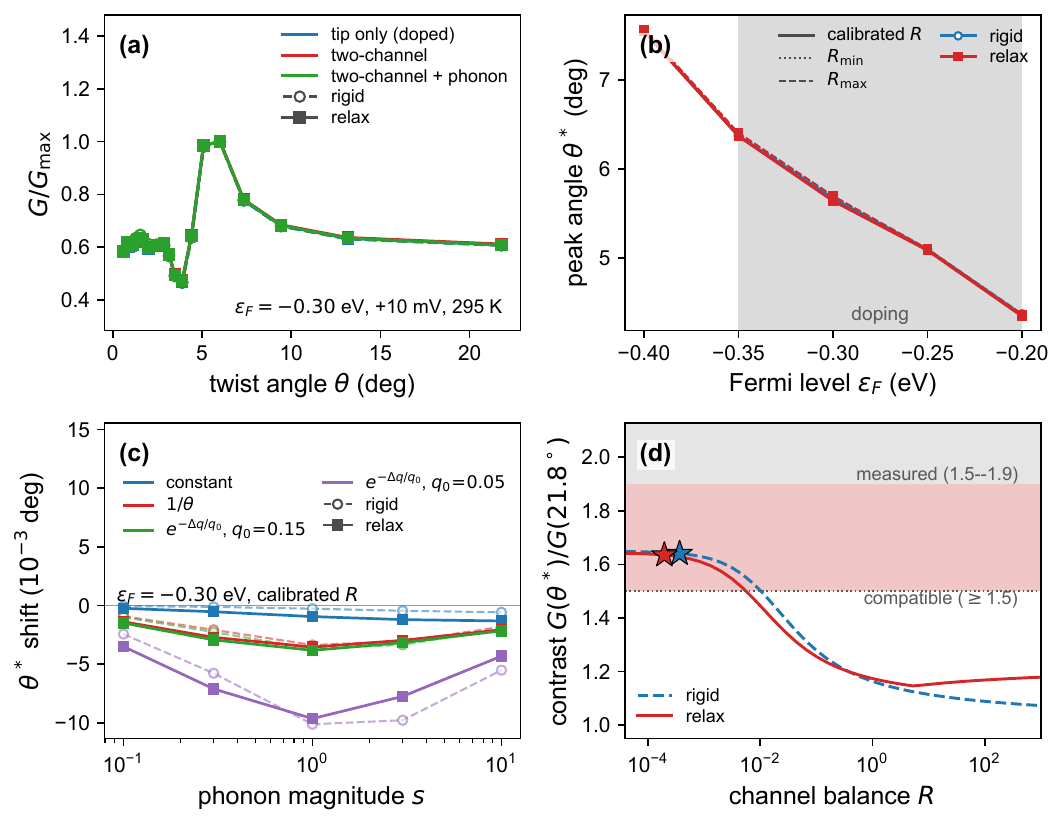}
\caption{
Unified transport model combining the doped finite-temperature tip term in series with the interlayer and phonon-assisted channels.
(a) Cell-averaged conductance versus twist angle at $\varepsilon_F=-0.30$~eV, $V=+10$~mV and $T=295$~K for the doped tip term alone, for the two-channel expression at the calibrated channel balance, and for the two-channel model including a phonon channel with magnitude $s=1$ and $1/\theta$ angular dependence.
The three curves are indistinguishable on this scale (open symbols: rigid; filled symbols: relaxed).
(b) Conductance peak angle $\theta^{*}$ of the combined conductance versus $\varepsilon_F$ at the calibrated balance at the extreme balances of the scan: the doping range corresponding to Fig.~\ref{fig:finiteT} is highlighted by the shaded region.
(c) Shift of $\theta^{*}$ relative to the tip-only calculation (in units of $10^{-3}$ degrees) versus the magnitude of the phonon channel $s$, for the four angular dependence profiles described in Appendix~\ref{app:phonon} ($q_0$ in \AA$^{-1}$), at $\varepsilon_F=-0.30$~eV and the calibrated channel balance.
(d) Angular contrast $G(\theta^{*})/G(21.8^{\circ})$ versus the channel balance $R$ (global scale of $n_j$); stars mark the calibrated balance, the light band indicates the range compatible with the experiment (contrast $\geq1.5$), and the darker band the measured contrast, $1.5$--$1.9$~\cite{Zhang2020}.}
\label{fig:combined}
\end{figure*}

Finally, the three transport mechanisms analyzed separately in this work, the doped finite-temperature tip term of Eq.~\eqref{Eq:finiteT-current}, the interlayer channel of Eq.~\eqref{Eq:acq}, and the phonon-assisted channel of Appendix~\ref{app:phonon}, are unified into a single, comprehensive calculation, whose results are shown in Fig.~\ref{fig:combined}.
The calculation proceeds in three steps.

(i) \emph{Doped, finite-temperature surface densities.}
For every atom $j$ in the top layer, the effective surface density entering the tip channel is given by
\begin{align}
\label{Eq:kernel-density}
n_{j}(\varepsilon_F,V,T)&=\int d\varepsilon\,\rho_{j}(\varepsilon)\,K(\varepsilon),\nonumber\\
K(\varepsilon)&=\frac{f(\varepsilon-\varepsilon_F-eV)-f(\varepsilon-\varepsilon_F)}{eV},
\end{align}
where $K(\varepsilon)$ is the thermal weight function of the junction, $f$ is the Fermi distribution at $T=295$~K, $\varepsilon_F$ is measured from the charge neutrality point. 
Since $\int K\,d\varepsilon=1$, $K$ reduces at $T=0$ to a rectangular box of width $e|V|$ between $\varepsilon_F$ and $\varepsilon_F+eV$, ensuring that $n_j>0$ for both polarities.
The energy-resolved LDOS $\rho_{j}(\varepsilon)$ of atom $j$, whose integral over a fixed energy window yields the density $\rho_j$ of Eq.~\eqref{Eq:acq-channels}, is obtained by direct diagonalization of the tight-binding Hamiltonian described in Sec.~\ref{sec:electronic-calculations} on symmetry-reduced $\Gamma$-centered $N_k\times N_k$ Monkhorst--Pack meshes.
Here, $N_k$ scales down from $501$ at $\theta=21.8^{\circ}$ to $27$ at $\theta=0.60^{\circ}$ to maintain a constant-reciprocal-space spacing $\Delta k=2\pi/(N_{k}L_{m}) \approx 1 \times 10^{-3}$~\AA$^{-1}$ across all moir\'e periods $L_m$.
This corresponds to an energy resolution of $\hbar v_F\Delta k\approx7$~meV (assuming $\hbar v_F=6.6$~eV\,\AA{} for graphene); a Lorentzian broadening of $\eta=5$~meV is analytically convolved with $K$.
Cells containing more than $6000$ atoms are diagonalized with a shift-and-invert sparse eigensolver restricted to the window $[-0.75,+0.35]$~eV around charge neutrality, which contains all the states with non-negligible weight in $K$.
The effective densities are evaluated for $\varepsilon_F=-0.40$, $-0.35$, $-0.30$, $-0.25$, and $-0.20$~eV and for $V=\pm10$ and $\pm50$~mV, on all 44 rigid and relaxed structures of this work.

(ii) \emph{Series combination.}
The densities $n_j$, expressed in units of states/eV (since $\int K\,d\varepsilon=1$), replace the window-integrated LDOS $\rho_j$ in Eq.~\eqref{Eq:acq-channels}. 
The phonon channel from Appendix~\ref{app:phonon} is then combined in parallel with  the interlayer term, yielding
\begin{align}
\label{Eq:combined}
G(\theta;\varepsilon_F,V,T)=\Big\langle\sum_{j\in{\rm top}}\frac{g_{j}\,\tilde S_{j}}{g_{j}+\tilde S_{j}}\Big\rangle_{\rm cell}
\end{align}
where
\begin{align}
g_{j}=R\,n_{j}\,e^{-3.91\,d_{j\text{-tip}}+10} \quad \mbox{and} \quad
\tilde S_{j}&=G^{\rm int}_{j}+p(\theta).
\end{align}
Here, the total conductance is averaged over a $40\times40$ grid of tip positions within the moir\'e cell, with the tip positioned $1$~\AA{} above the local height of the top layer and periodic images included up to a cutoff distance of $10$~\AA.
The factor $R$ (not to be confused with the interatomic distance entering the hopping amplitudes of Sec.~\ref{sec:electronic-calculations}) absorbs the normalization of $n_j$ relative to the empirical prefactors of Eq.~\eqref{Eq:acq-channels} and determines the balance between the two channels [Fig.~\ref{fig:acq}(d)]. 
In the limit $R\to0$, the junction is entirely tip limited, $G\to\langle\sum_j g_j\rangle$. 
Conversely, for $R\to\infty$, it becomes interlayer limited, $G\to\langle\sum_j\tilde S_j\rangle$, and all information regarding the surface LDOS is lost.

We scan $R$ over four orders of magnitude and calibrate it separately for the rigid and relaxed structure families such that the angular contrast $G(\theta^{*})/G(21.8^{\circ})$ of the two-channel curve at $\varepsilon_F=-0.30$~eV and $V=+10$~mV matches the experimental range of $1.5$--$1.9$ Ref.~\cite{Zhang2020} as closely as possible. 
The calibrated values yield a contrast of $1.64$ for the rigid structures and $1.63$ for the relaxed ones, representing the maximum attainable contrast at this doping level.
The phonon channel is scanned as detailed in Appendix~\ref{app:phonon}, taking $p(\theta)=s\,\bar S\,u(\theta)$, where $\bar S$ is the elastic interlayer conductance $G^{\rm int}_{j}$ averaged over atoms and structures, $s=0.1$--$10$ (five values) sets the magnitude of the phonon channel with respect to $\bar S$, and $u(\theta)$ is one of the four angular shapes defined there, normalized to unity at $\theta=5.09^{\circ}$.

(iii) \emph{Peak angle.}
For each of the $20$ combinations $(\varepsilon_F,V)$ of step (i), $5$ channel balances, $21$ phonon settings (no phonon channel, plus five magnitudes for each of the four shapes), and $2$ structural families, the peak angle $\theta^{*}$ is 
determined by fitting a parabola to the three sampled angles surrounding the discrete maximum and extracting its vertex. 
This value is directly compared with that of the tip-only doped calculation ($R\to0$, $p=0$) for the same $(\varepsilon_F,V)$, which is the quantity shown in Fig.~\ref{fig:finiteT}.
A combination is considered compatible with the experiment when the angular contrast of its two-channel curve meets or exceeds $1.5$.

The results are summarized in Fig.~\ref{fig:combined}.
For every compatible channel balance, the maximum of the combined conductance coincides with that of the doped tip-only calculation to within $0.2^{\circ}$, and to within $0.01^{\circ}$ at the calibrated balance. 
The inclusion of the interlayer and phonon channels serves only to reduce the contrast, without displacing the peak position. 
Furthermore, rigid and relaxed structures agree to within $0.03^{\circ}$.
Within the range $-0.35\le\varepsilon_F\le-0.20$~eV shown in Fig.~\ref{fig:finiteT}(d) the combined $\theta^{*}$ lies between $4.2^{\circ}$ and $6.5^{\circ}$ for $V=\pm10$~mV, with the two polarities agreeing to within $0.35^{\circ}$.
Displacements larger than $1^{\circ}$ appear only in the interlayer-limited regime, whose nearly flat angular response (contrast below $1.5$) is ruled out by the experiment.

\section{In-plane versus out-of-plane relaxation}
\label{app:ipop}

To disentangle the roles of the in-plane and out-of-plane relaxation components, we construct two families of constrained (hybrid) structures for $\theta = 5.09^{\circ}$, $2.45^{\circ}$, and $1.12^{\circ}$:
(i) in-plane relaxation only (relaxed $xy$ coordinates, flat layers at the rigid spacing) and (ii) out-of-plane relaxation only (rigid $xy$ coordinates, relaxed $z$ corrugation).
For this comparison all four structures of each unit cell (rigid, in-plane only, out-of-plane only, and fully relaxed) are computed with an interlayer hopping that depends explicitly on the actual local interlayer distance, allowing the corrugation to act directly on the electronic structure.

\begin{figure}[t!]
\centering
\includegraphics[width=0.9\columnwidth]{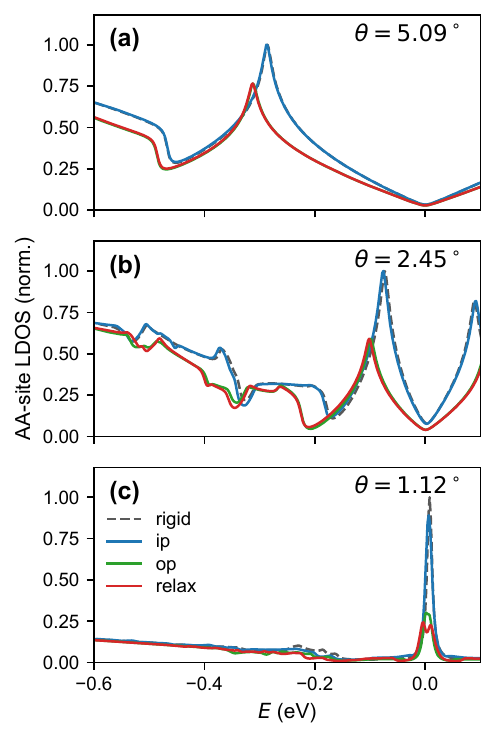}
\caption{
In-plane versus out-of-plane relaxation: LDOS at the AA site for the four structure types at (a) $\theta=5.09^{\circ}$, (b) $\theta= 2.45^{\circ}$, and (c) $\theta= 1.12^{\circ}$. 
The in-plane-only curves lie on top of the rigid ones, whereas the out-of-plane-only curves closely match the fully relaxed ones, in which the van Hove singularities are slightly shifted to lower energies and reduced in intensity.}
\label{fig:ipop}
\end{figure}

Figure~\ref{fig:ipop} shows the LDOS at the AA site for the three angles.
The in-plane-only spectra are nearly indistinguishable from the rigid ones, reflecting the small magnitude of the in-plane displacements at these angles (below $0.03$~\AA{} for $\theta\ge 2.45^{\circ}$).
The visible changes are instead carried by the $\approx 0.15$~\AA{} out-of-plane corrugation, which locally weakens the interlayer hopping in the bulged AA regions: the van Hove singularities shifts slightly to lower energies and lose intensity, causing the out-of-plane-only spectra closely lie on top of the fully relaxed ones.
The corresponding effect on the contact current under the experimental conditions is a mild, smooth amplitude modulation, dominated by the out-of-plane component and purely electronic in origin (persisting when the tip-distance weighting is removed). 
It should be stressed, the position of the conductance maximum remains almost unchanged, with all four structure types sharing the same peak angle.
Thus, neither relaxation component creates or shifts the conductance maximum, fully supporting the conclusions of the main text.

\section{Bound on the phonon-assisted channel}
\label{app:phonon}

Phonon-assisted processes provide an additional interlayer conduction channel, in parallel with the elastic (geometric) one, by supplying the momentum mismatch $\Delta q(\theta) = 2 |{\bf K}|\sin(\theta/2)$
, with $|{\bf K}|=4\pi/(3a)=1.70$~\AA$^{-1}$, between the rotated Dirac cones~\cite{Perebeinos2012, Perebeinos2024}. 
These results give a phonon-mediated conductance that varies smoothly with the twist angle at room temperature, dominated by the interlayer breathing (layer-breathing, ZO$'$) mode, a behavior consistent with the large-angle measurements of Ref.~\cite{Yu2020}.

In the junction expression of Appendix~\ref{app:acq}, the phonon channel enters as $G^{\rm int}_{j} \rightarrow G^{\rm int}_{j} + p(\theta)$, with $p$ spatially uniform.
To bound its possible effect, we scan $p(\theta) = s\, \bar{S}\, u(\theta)$ over magnitudes $s = 0.1$--$10$ relative to the elastic interlayer channel $\bar S$, and over monotonic shapes $u(\theta)$ ranging from constant to $1/\theta$ and to exponential decay in the phonon momentum, $u = e^{-\Delta q(\theta)/q_0}$ with $q_0 = 0.15$ and $0.05$~\AA$^{-1}$, recomputing the cell-averaged series conductance at every twist angle.
These functional forms are chosen to bound the angular dependence predicted in Ref.~\cite{Perebeinos2012}, $G_{\rm ph}\propto n_{\theta}/\omega_{\theta}$ [Eq.~(4) therein], where $\omega_{\theta}$ is the frequency of the layer-breathing mode at the momentum mismatch $\Delta q(\theta)$ and $n_{\theta}$ its thermal population.
At room temperature, this theoretical dependence is smooth, monotonic, and varies by less than an order of magnitude across the relevant angular range (excluding the limit of very small angles). 
In contrast, between $1^{\circ}$ and $21.8^{\circ}$ the constant, $1/\theta$ and exponential shapes vary by factors of $1$, $22$, $60$ and $2\times10^{5}$, respectively. 
The family of shapes scanned here thus brackets the predicted behavior comfortably on both sides.

A $\theta$-independent phonon channel of any magnitude leaves the position of the maximum strictly unchanged.
Angle-dependent shapes displace it by at most $0.7^{\circ}$ (our angular sampling resolution) in any channel balance compatible with the observed angular contrast of the current: for balances yielding a peak-to-large-angle contrast 
$\geq 1.5$, 
as observed experimentally ($1.5$--$1.9$ in Ref.~\cite{Zhang2020}) and obtained in Fig.~\ref{fig:finiteT} ($1.6$--$2.0$), the maximum shift is $0.7^{\circ}$ for every shape and magnitude considered.
Larger displacements occur only in a strongly interlayer-limited regime whose nearly flat angular response (contrast $\lesssim 1.4$) is itself incompatible with the measured maximum.

 {Finally, the corrugation and reconstruction below $\approx5^{\circ}$ renormalize the relevant low-energy modes by at most a few tens of cm$^{-1}$~\cite{Gadelha2021,Koshino2019moire}, yielding a smooth modulation of the phonon channel well within the family of magnitudes and shapes scanned above.
We conclude that the phonon-assisted channel can neither produce the conductance maximum nor displace it appreciably. 
A full electron-phonon calculation on relaxed moir\'e cells containing $10^{3}$--$10^{4}$ atoms is beyond the scope of this work and, given the bound above, unnecessary for the present conclusions.}

\section{Numerical details}
\label{app:numerics}

 {
Rather than employing a terminator in the continued-fraction expansion of the HHK recursion method, convergence is  achieved by systematically increasing both the recursion depth and the real-space cluster size for each twist angle until the results fully stabilize. 
The number of recursion steps ranges from about $10^{3}$, for the largest twist angles, to about $5\times10^{3}$, for the smallest ones, with cluster sizes ranging from about $8\times10^{6}$ to $10^{8}$ atomic sites, respectively, larger moir\'e cells and deeper recursion being necessary as $\theta$ decreases to resolve the van Hove singularities approaching the Fermi level. 
Throughout these calculations, an imaginary energy broadening of $\eta=0.1$~meV is used.

 {
For each twist angle, the cluster size and recursion depth are increased until the LDOS at the relevant sites no longer changes appreciably. 
Where direct diagonalization remains computationally tractable, it matches the HHK results. 
This is consistent with the extensive validation of the same HHK implementation reported in our previous work~\cite{Vidarte2022,Vidarte2024Q,Vidarte2025}, where it was benchmarked  against direct diagonalization, independent analytical and reciprocal-space calculations, and experimental spectroscopy. 
Regarding the sensitivity to $\eta$: in the continued-fraction recursion formalism the energy of a resolved spectral feature is determined by the recursion coefficients (i.e., by $\theta$ and the converged depth), not by $\eta$, which only sets its numerical width. 
Since $\eta$ is three orders of magnitude smaller than the narrowest integration window ($100$~meV) and well below the characteristic width of the van Hove singularities, whether a singularity falls inside a given window, and hence the position of the conductance maximum, is controlled by the system's electronic structure, not by the spectral broadening.
}

Conversely, the cell-averaged and per-atom densities reported in Figs.~\ref{fig:finiteT}--\ref{fig:combined} and Appendices~\ref{app:acq}--\ref{app:phonon} are computed  by direct diagonalization on the Monkhorst--Pack meshes specified in Appendix~\ref{app:acq}, with a Lorentzian broadening of $5$~meV, well below the thermal width $k_BT=25$~meV entering Eq.~\eqref{Eq:finiteT-current}.

\section{Contact-induced Fermi-level shift under the tip}
\label{app:doping}
To estimate the local Fermi-level position under a metal-coated probe we solve the contact electrostatics of Refs.~\cite{Giovannetti2008,Khomyakov2009}.
In this framework, the charge transferred across the metal--graphene interface is determined by the work-function difference $W_M-W_G$, by the interface-dipole step $\Delta_c(d)$ and by the capacitance of the contact, 
\begin{equation}
\varepsilon_F = -\left[W_M-W_G-\Delta_c(d)-\alpha\, n(\varepsilon_F)\,(d-d_s)\right],
\end{equation} 
where $d$ is the metal--graphene separation, $n(\varepsilon_F)$ is the number of carriers transferred per graphene unit cell of area $A_{\rm c}=5.18$~\AA$^{2}$, $\alpha=e^{2}/(\epsilon_0 A_{\rm c})=34.9$~eV/\AA{} ($\epsilon_0$ being the vacuum permittivity), $d_s=2.4$~\AA{} the screening-plane offset, and $\Delta_c(d)$ follows the parametrization of  Ref.~\cite{Khomyakov2009}.
Instead of assuming the linear DOS of single-layer graphene, $n(\varepsilon_F)$ is obtained by integrating the actual density of states of each twisted bilayer computed in this work from charge neutrality to $\varepsilon_F$ (including both layers, normalized to one state per $p_z$ orbital and referred to the single-layer unit cell). 
When evaluated using the linear DOS, the model reproduces the Fermi-level shifts tabulated in Ref.~\cite{Khomyakov2009} for Al, Ag, Cu, Au and Pt to within $0.1$~eV.
Figure~\ref{fig:doping}(a) shows $\varepsilon_F(\theta)$ for $W_M=5.55$~eV, $d=3.3$~\AA{} and $\Delta_c=0.9$~eV, together with the range spanned by $W_M\in[5.3,5.8]$~eV, $d\in[3.0,3.4]$~\AA{} and $\Delta_c\in[0.8,1.0]$~eV.
The Fermi-level shift varies by less than $70$~meV over the entire angular range and is virtually insensitive to structural relaxation, thereby justifying the use of constant, angle-independent $\varepsilon_F$.
Figure~\ref{fig:doping}(b) shows the shift as a function of the coating work function at $\theta=5.09^{\circ}$: the doping sign changes at $W_M=W_G+\Delta_c\approx5.4$~eV, falling within the range reported for Ir, whereas Pt and Pt-rich coatings give $p$-type doping of $0.2$--$0.4$~eV~\cite{Michaelson1977,Derry1989}.

\begin{figure*}[t]
\centering
\includegraphics[width=0.9\textwidth]{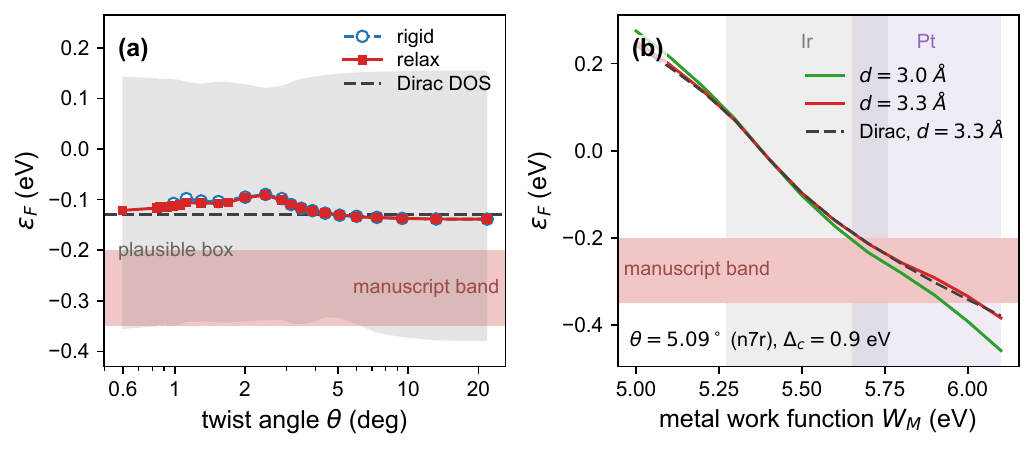}
\caption{
Contact-induced Fermi-level shift under a metal-coated tip obtained from the self-consistent contact model of Refs.~\cite{Giovannetti2008,Khomyakov2009} using the density of states of the twisted bilayer. (a) $\varepsilon_F$ versus twist angle for $W_M=5.55$~eV, $d=3.3$~\AA{} and $\Delta_c=0.9$~eV (open symbols: rigid; filled symbols: relaxed).
The dashed line represents the result for the linear (Dirac) density of states of monolayer graphene.
The grey band spans $W_M\in[5.3,5.8]$~eV, $d\in[3.0,3.4]$~\AA{} and $\Delta_c\in[0.8,1.0]$~eV and the shaded horizontal band indicates the range used in Fig.~\ref{fig:finiteT}. 
(b) $\varepsilon_F$ versus the work function of the coating at $\theta=5.09^{\circ}$ for $d=3.0$ and $3.3$~\AA{}. 
The vertical shaded regions mark the reported work-function ranges of Ir and Pt.}
\label{fig:doping}
\end{figure*}

\begin{acknowledgments}
    This work was supported by the Brazilian funding agencies CNPq, CAPES, FAPERJ, FAPESP, and FAPEMIG. 
    ESM acknowledges financial support from ANID project 1221301.
\end{acknowledgments}

\bibliography{TBG,methods}

\end{document}